\documentclass[
journal=jpcbfk,
manuscript=article,
layout=traditional,
chaptertitle=true]{achemso}

\usepackage{amssymb}
\usepackage{amsmath}
\usepackage{graphicx}
\usepackage{xcolor}
\usepackage{makecell}
\usepackage{bm}
\usepackage{currvita}
\usepackage{float}
\usepackage[normalem]{ulem}

\title{Computational Design of Molecular Probes for Electronic Pre-Resonance Raman Scattering Microscopy}

\author{Jiajun Du}
\altaffiliation{Contributed equally to this work}
\author{Xuecheng Tao}
\altaffiliation{Contributed equally to this work}
\author{Tomislav Begu\v{s}i\'c}
\email{tbegusic@caltech.edu}
\author{Lu Wei}
\email{lwei@caltech.edu}
\affiliation{Division of Chemistry and Chemical Engineering, California Institute of Technology, Pasadena, California 91125, USA}

\begin{document}

\begin{tocentry}
    \centering 
    \includegraphics[height=4.45cm]{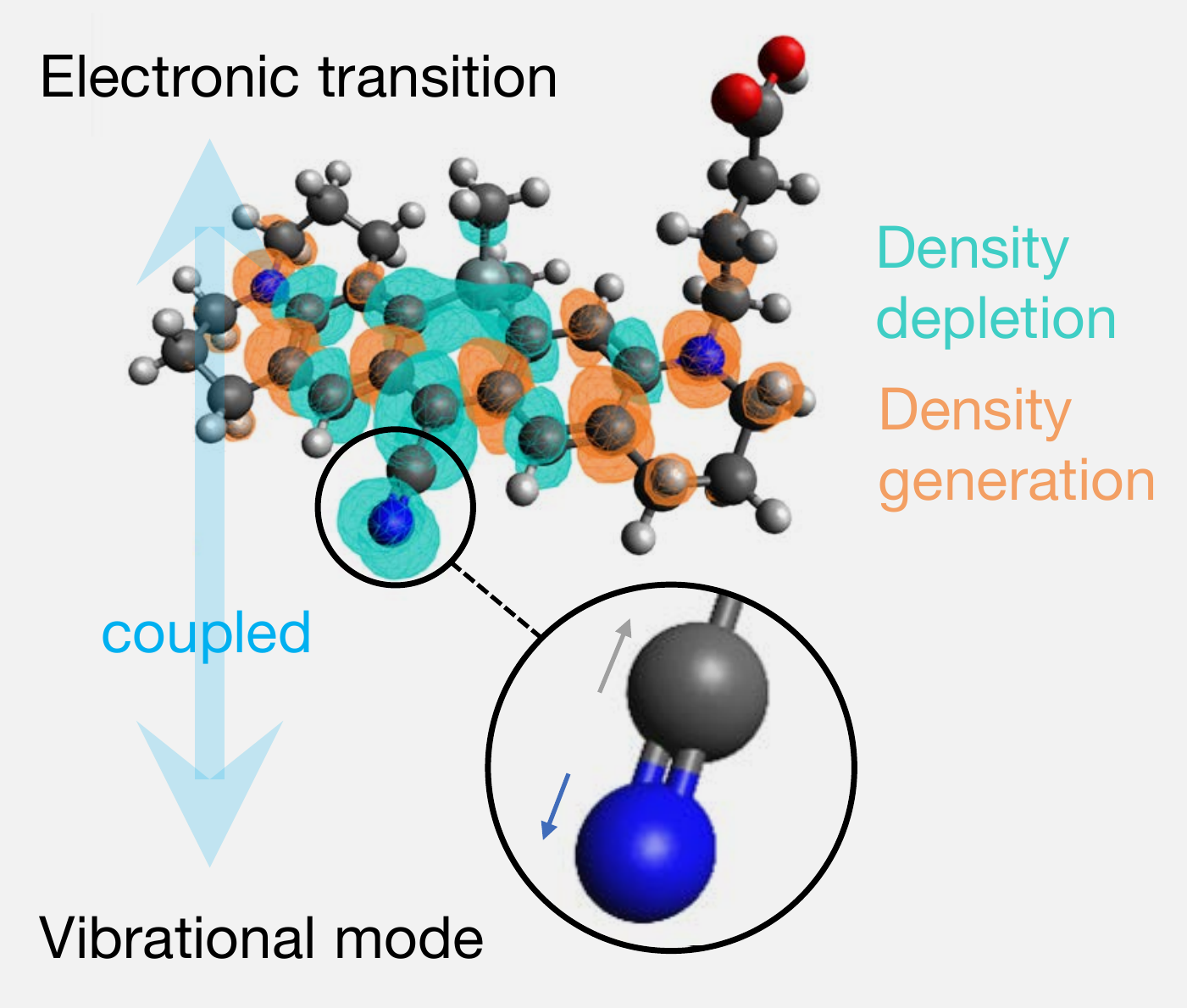}
\end{tocentry}

\begin{abstract}
Recently developed electronic pre-resonance stimulated Raman scattering (epr-SRS) microscopy, in which the Raman signal of a dye is significantly boosted by setting the incident laser frequency near the electronic excitation energy, has pushed the sensitivity of SRS microscopy close to that offered by confocal fluorescence microscopy. 
Prominently, the maintained narrow line-width of epr-SRS also offers high multiplexity that breaks the “color barrier” in optical microscopy. 
However, detailed understandings of the fundamental mechanism in these epr-SRS dyes still remain elusive.
Here, we combine experiments with theoretical modeling to investigate the structure-signal relationship,
aiming to facilitate the design of new probes and expanding epr-SRS palettes. Our ab initio approach employing the displaced harmonic oscillator (DHO) model provides a consistent agreement between simulated and experimental SRS intensities of various triple-bond bearing epr-SRS probes with distinct scaffolds. We further review two popular approximate expressions for epr-SRS, namely the short-time and Albrecht A-term equations, and compare them to the DHO model. Overall, the theory allows us to illustrate how the observed intensity differences between molecular scaffolds stem from the coupling strength between the electronic excitation and the targeted vibrational mode, leading to a general design strategy for highly sensitive next-generation vibrational imaging probes.
\end{abstract}

\section{Introduction}
Over the last one and half decades, stimulated Raman scattering (SRS) microscopy has emerged as an important vibrational bio-imaging modality complementary to standard fluorescence microscopy. Although SRS has enhanced the otherwise weak spontaneous Raman transition by up to 10$^8$-fold through stimulated emission amplification\cite{ploetz2007femtosecond,freudiger2008label,ozeki2009analysis}, 
its current sensitivity of non-resonant probes is still largely limited to micromolar to millimolar range\cite{du2022bringing},
restricting probing the rich chemical information of dilute biomolecules {\it in vivo},
which is usually in the nanomolar to low micromolar range.
This sensitivity gap has proven to be successfully tackled by customized Raman probes\cite{du2022bringing}. 
Among numerous Raman probes, some of the most sensitive ones up to date are the pyronin-based electronic pre-resonance (epr) enhanced Manhattan Raman scattering (MARS) dyes\cite{wei2017super}. When the pump wavelength is tuned to be close to the electronic excitation energy, i.e. under the epr condition, the vibrational mode coupled to the electronic state would be selectively amplified with enhanced SRS signals.
By carefully tuning the absorption of the dyes (660--790 nm) to moderately close to the laser wavelength (800$\sim$900 nm), 
SRS intensities of the triple bonds (nitriles or alkynes) when conjugated into the conjugation systems of these dyes have been found to be pre-resonantly enhanced by up to 10$^4$ folds (detection limit down to 250 nM) with a well-maintained high signal-to-background ratio. Since the invention of MARS dyes, numerous exciting imaging applications have been achieved. 
The MARS dyes enable super-multiplexed 
($>$20 channels) vibrational imaging by taking advantage of the narrow linewidth of Raman peaks (peak width about 10 cm$^{-1}$, $\sim$50--100 times narrower than fluorescent peaks) from triple bonds in the cell-silent region (1800--2800 cm$^{-1}$) where there are no background signals from endogenous molecules.\cite{wei2017super, miao20219, shi2022highly}
These ideas inspired the design of multi-functional Raman probes including photo-switchable, photo-activatable, and turn-on enzymatic probes\cite{lee2021toward, ao2021switchable, shou2021photoswitchable, fujioka2020multicolor, kawatani20229}. 
They also paved the way for all-far-field single-molecule Raman spectroscopy and imaging without plasmonic enhancement by stimulated Raman excited fluorescence (SREF)\cite{xiong2019stimulated}. Thus the superb vibrational properties and versatility of epr MARS dyes are widely recognized, making them essential for a wide range of vibrational spectroscopy and imaging. 
However, these dyes are still the only set of triple-bond bearing epr-SRS dyes until now and the principle of designing such strong Raman probes is still inconclusive. This has largely restricted the development of new epr-SRS scaffolds to further increase the sensitivity and expand the multiplexity, a central topic in the current development of SRS imaging. While the Albrecht A-term pre-resonance approximation equation was previously adopted to fit the dependence of epr-SRS signals with a single parameter of laser detuning, the treatment ignored the structure dependent factors as it assumes all frequency-independent factors as a constant. \cite{wei2017super}
The necessity to rationally explore and design new epr-SRS scaffolds hence sets a high demand for a more systematic theory to understand and predict the dependence of the  epr-SRS signals on molecule structures (i.e., a structure-intensity relationship). 

\begin{figure}[!t]
    \centering \includegraphics[width=\linewidth]{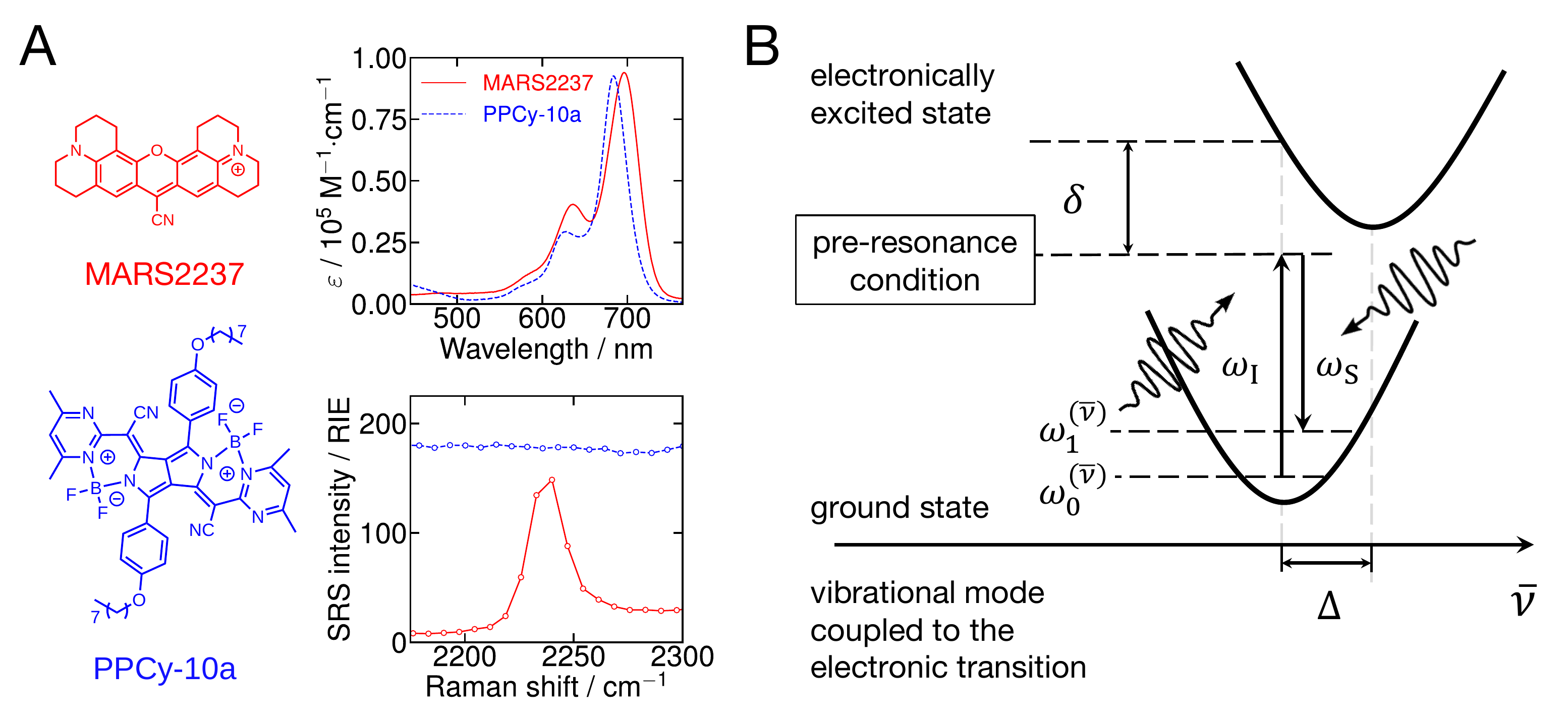}
    \caption{\label{scheme} 
    {\bf (A)}  Absorption and SRS spectra of MARS2237 (red) and PPCy-10a (blue). 
    SRS intensities are reported relative to EdU (RIE) values.
    $\varepsilon$ stands for the molar extinction coefficient.
    {\bf (B)} Scheme of the electronic pre-resonance stimulated Raman scattering  (epr-SRS) process, in which the incident (pump) light with frequency $\omega_{\rm I}$ excites the vibrational mode $\bar{\nu}$ from its ground to the first excited vibrational state.
    The pump pulse is detuned by $\delta$ from the vertical excitation frequency between the ground and first excited electronic states of the molecule. The scattering process is stimulated by a Stokes (probe) pulse of frequency $\omega_{\rm S}$ and proceeds through a coupling between the vibrational and electronic degrees of freedom, controlled by the dimensionless displacement parameter $\Delta$.
    }
\end{figure}

Indeed, our initial screening of molecular candidates for new epr-SRS probes revealed that the structure related factors play a crucial role. 
For example, a series of pyrrolopyrrole cyanine (PPCy) dyes were originally identified by us to be promising candidates of epr-SRS probes. 
PPCy dyes are neutral and have two nitrile groups in the conjugation system with adjustable absorption between 680 and 800 nm\cite{fischer2009pyrrolopyrrole}. The absorption spectrum of one of those dyes, namely the PPCy-10a molecule (Figure~\ref{scheme}A, blue), almost overlaps with that of MARS2237 (Figure~\ref{scheme}A, red), one of the well-validated MARS dyes.
The absorption maximum and molar extinction coefficient of these two molecules are also very close, implying that a similar epr-SRS signal should be expected based on the detuning and oscillator strength dependence implied from the Albrecht A-term pre-resonance approximation equation \cite{wei2017super}.
However, as a stark difference to MARS2237, which presented a clear and sharp epr-SRS peak, we barely see any epr-SRS signal from the nitrile groups of PPCy-10a under the same measurement conditions (Figure~\ref{scheme}A). 
This observation points out that it is not effective enough to identify novel epr-SRS probes only through the experimentally measured absorption quantities. Instead, it indicates that the epr-SRS process relies heavily on the specific molecular structures. To decipher the structure-dependent factors underlying this vibronic process, we turn to more accurate quantum chemistry approaches.

\section{Theory}

Theoretical tools that can accurately and reliably predict Raman intensity could greatly help us understand the key factors involved in epr-SRS. Although established computational methods for simulating resonance and pre-resonance Raman spectra exist,\cite{Kumble1998, bailey2000interference, petrenko2007analysis, Petrenko2012, Egidi2014, Baiardi2014, Rao2016, Quincy2018, de2019efficient, Mattiat2021} they have been largely used for simulating the full spectra of individual molecules. Here, we explore an alternative task, in which we compare the intensities of a single peak, corresponding to the nitrile (C$\equiv$N) or alkyne (C$\equiv$C) bond stretch, in the pre-resonance Raman spectra of multiple large molecules (50--100 atoms). Theoretically, the SRS signal intensity is\cite{tannor2006introduction, schatz2002quantum}
\begin{equation}
\label{eq:crosssec}
    I_{0 \to 1}^{(\bar{\nu})} \propto \sum_{\rho \lambda} |\alpha^{(\bar{\nu})}_{0 \to 1}(\omega_{\rm I} )_{\rho \lambda}|^2,
\end{equation}
where $0$ and $1$ denote the ground and first excited vibrational states of mode $\bar{\nu}$ in the electronic ground state, $\alpha^{(\bar{\nu})}_{0 \to 1}(\omega_{\rm I})$ is the frequency-dependent polarizability matrix element between these vibrational states, $\omega_{\rm{I}}$ is the incident light frequency, and $\rho$ and $\lambda$ are the polarizations of the incident and scattered lights. In eq~\ref{eq:crosssec}, we omit the prefactor that depends on the incident light (pump) frequency and the Stokes pulse intensity because these factors cancel out when the SRS intensity is reported with respect to a standard reference.
The conventional frequency-domain approach formulates the polarizability through the Kramers-Heisenberg-Dirac (KHD) equation\cite{kramers1925streuung, dirac1927quantum} 
\begin{equation}
\label{eq:KHD}
    \alpha^{(\bar{\nu})}_{0 \to 1}(\omega_{\rm{I}})_{\rho \lambda} = - \frac{1}{\hbar}\sum_n \left[ \frac{\langle \psi_1^{(\bar{\nu})} | \hat{\mu}_{\rho} | \psi^{\prime}_n \rangle\langle \psi^{\prime}_n | \hat{\mu}_{\lambda} | \psi_0^{(\bar{\nu})} \rangle}{\omega_{\rm{I}} - \omega^{\prime}_n + \omega_0^{(\bar{\nu})} + i \gamma} - \frac{\langle \psi_1^{(\bar{\nu})} | \hat{\mu}_{\lambda} | \psi^{\prime}_n \rangle\langle \psi^{\prime}_n | \hat{\mu}_{\rho} | \psi_0^{(\bar{\nu})} \rangle}{\omega_{\rm{I}} + \omega^{\prime}_n - \omega_1^{(\bar{\nu})} - i \gamma}\right],
\end{equation}
which involves a sum over all vibrational states $\psi^{\prime}_n$ (with energies $\hbar \omega^{\prime}_n$) of the excited electronic state that is near resonance with the incident light.
$\hbar \omega^{(\bar{\nu})}_0$ and $\hbar \omega^{(\bar{\nu})}_1$ correspond to the energies of the ground and first excited vibrational states of the mode of interest ($\bar{\nu}$), respectively.
$\hat{\mu}_\rho = \hat{\vec{\mu}} \cdot \vec{\varepsilon}_{\rm S}$, $\hat{\mu}_\lambda= \hat{\vec{\mu}} \cdot \vec{\varepsilon}_{\rm I}$ are the projections of transition dipole moment $\hat{\vec{\mu}}$ along the scattered and incident light polarizations, respectively, and $\gamma$ is the dephasing parameter. In the remainder, we will assume the Condon approximation, $\hat{\vec{\mu}} \equiv \vec{\mu}(\hat{q}) \approx \vec{\mu}(q_{\rm{eq}})$, in which the coordinate dependence of the transition dipole moment is neglected. $q_{\rm{eq}}$ denotes the ground-state equilibrium geometry.

Time-domain approach, popularized by Heller and Tannor,\cite{lee1979time, heller1982simple} offers an efficient alternative to evaluating the above sum-over-states formula.  Here, the frequency-dependent polarizability is written as the half-Fourier transform
\begin{align}
\label{eq:alpha_time}
    \alpha^{(\bar{\nu})}_{0 \to 1}(\omega_{\rm{I}})_{\rho \lambda} = \frac{i}{\hbar} \mu_{\lambda}\mu_{\rho} \left[\int_0^{\infty} C(t) \Gamma(t) \hspace{4pt} 
    e^{i \omega_{\rm{I}} t} dt + \int_0^{\infty} C(t) \Gamma(t) \hspace{4pt} 
    e^{-i (\omega_{\rm{I}} - \omega^{(\bar{\nu})}) t} dt \right]
\end{align}
of the time correlation function
\begin{equation}
\label{eq:C_t}
    C(t) = \langle \psi^{(\bar{\nu})}_1 | e^{-i \hat{H}_\textrm{e} t / \hbar} | \psi^{(\bar{\nu})}_0 \rangle e^{i \omega_0^{(\bar{\nu})} t},
\end{equation}
where $\hat{H}_\textrm{e}$ is the excited-state Hamiltonian, and we denote $\omega^{(\bar{\nu})} =  \omega^{(\bar{\nu})}_1 - \omega^{(\bar{\nu})}_0$ for simplicity. $\Gamma(t)$ is the dephasing term in its more general form; for example, $\Gamma(t) = e^{-\gamma t}$ corresponds to the Lorentzian lineshape in the KHD expression (eq~\ref{eq:KHD}).
In words, the computation of $C(t)$ requires the time propagation of a quantum wavepacket $|\psi(t)\rangle = \exp(-i \hat{H}_\textrm{e} t / \hbar) | \psi^{(\bar{\nu})}_0 \rangle$ in the excited electronic state, which is computationally costly for molecular Raman probes with a typical size of 50--100 atoms, even with, for example, trajectory-guided Gaussian wavepacket approaches.\cite{Rohrdanz_Cina:2006,Makhov_Shalashilin:2021,Worth_Lasorne:2020,Conte_Ceotto:2020,Bonfanti_Pollak:2018,Werther_Grossmann:2020,Werther_Grossmann:2021,Curchod_Martinez:2018,Prlj_Vanicek:2020,Vanicek_Begusic:2021,Begusic_Vanicek:2021a,Begusic_Vanicek:2022} In fact, the number of ab initio computations in the excited electronic state should be minimized to allow for efficient analysis of relatively large molecules. 

The displaced harmonic oscillator (DHO) model\cite{Petrenko2012} offers a practical way to approximate the time propagation in the excited state.
Within this model, it is assumed that the ground and excited potential energy surfaces can be sufficiently accurately represented as two harmonic potentials with equal force constants (with frequencies $\omega^{(\nu)}$) but different minima. 
Note that $\nu$ represents an arbitrary mode, and is distinguished from $\bar{\nu}$, the mode of interest that gives the Raman signal.
In this case, the time correlation function of eq~\ref{eq:C_t} simplifies into 
\begin{equation}
\label{eq:DHO}
    C^{\rm{DHO}}(t) = - \frac{\Delta_{\bar{\nu}}}{\sqrt{2}} (1-e^{-i \omega^{(\bar{\nu})} t})
    \prod_{\nu} \left[e^{- (1 - i\omega^{(\nu)}t - e^{-i \omega^{(\nu)} t}) \Delta_{\nu}^2 / 2}\right] e^{- i (\delta + \omega_{\rm I}) t}.
\end{equation}
$\Delta_\nu$ is the dimensionless distance between the ground- and excited-state minima along mode $\nu$ and is expressed within the vertical gradient model\cite{Baiardi2014} as
\begin{align} \label{eq:delta_nu}
    \Delta_\nu = \frac1{\omega^{(\nu)}} \sqrt{\frac{f_\nu^2}{\hbar \omega^{(\nu)}}},
\end{align}
where $f_\nu$ is the gradient of the excited-state potential energy with respect to the mass-scaled normal mode $\nu$ (see Figure~\ref{scheme}B) evaluated at the ground-state equilibrium geometry. The dimensionless displacement parameters $\Delta_{\nu}$ are directly related to the well-known Huang--Rhys factors $S_{\nu}=\Delta_{\nu}^2/2$. In eq~\ref{eq:DHO}, $\delta = \omega_{\rm{eg}} - \omega_{\rm{I}}$ is the difference (detuning) between the vertical excitation frequency $\omega_{\rm{eg}}$ and the incident light frequency. Equation~\ref{eq:alpha_time} is combined with the approximations in eqs~\ref{eq:DHO} and \ref{eq:delta_nu} to provide the major simulation protocol for this study. In addition, because the laser frequency is sufficiently close to the vertical excitation gap, only the resonant part (first term on the right-hand side of eq~\ref{eq:alpha_time}) was computed.

Finally, we close the review of the methods by pointing out that two simple but useful expressions can be further derived from eq~\ref{eq:DHO} under the pre-resonance conditions. First, following Heller, Sundberg, and Tannor\cite{heller1982simple, tannor2006introduction}, a short-time expansion of the time correlation function $C^{\rm{DHO}}(t)$ leads to
\begin{align} \label{c_shorttime}
    C^{\rm{ST}}(t) = - i s_{\bar{\nu}} t e^{-s^2 t^2 / 2 - i(\delta + \omega_{\rm{I}}) t},
\end{align}
where $s_\nu = \Delta_\nu \omega^{(\nu)} / \sqrt{2}$ and $s^2 = \sum_\nu s_\nu^2$. Substituting eq~\ref{c_shorttime} into the first (resonant) term of eq~\ref{eq:alpha_time} yields the short-time expression for the SRS intensity
\begin{align}
    \label{alpha_shorttime}
    I_{0 \to 1}^{(\bar{\nu}), \rm ST} &\propto 
    \frac{\mu^4 s_{\bar{\nu}}^2}{\hbar^2 s^4}
    \left| \int_0^\infty t 
    e^{-t^2/2+i\delta t/s} dt \right|^2,
\end{align}
where we used $\sum_{\rho \lambda} \mu_\rho^2 \mu_\lambda^2 = \mu^4$ and introduced the dipole strength $\mu^2 = \sum_{\rho = x,y,z} \mu_{\rho}^2$.
Alternatively, the large detuning limit can be applied directly to the KHD formula in the frequency domain, which leads to the well-known Albrecht A-term equation but now with calculable structure-dependent factors\cite{Albrecht1961,asher1988uv} 
\begin{align} \label{albrecht}
    I_{0 \to 1}^{(\bar{\nu}), \text{Albrecht}} &\propto 
    \frac{4 \mu^4 s_{\bar{\nu}}^2}{\hbar^2} 
    \left[\frac{\omega_{\rm{I}}^2 + \omega_{\rm{eg}}^2}{(\omega_{\rm{I}}^2 - \omega_{\rm{eg}}^2)^2}\right]^2 \approx \frac{\mu^4 s_{\bar{\nu}}^2}{\hbar^2 \delta^4}.
\end{align}
The first part of eq~\ref{albrecht} corresponds to the most common form that includes both resonant and non-resonant terms, whereas the final right-hand side expression is an approximate form assuming $\delta \ll \omega_{\rm eg}$. Although more approximate than eq~\ref{eq:DHO}, the short-term and Albrecht's expressions provide additional insight into the origins of strong pre-resonance Raman signals. For example, Albrecht's expression reveals a strong ($1/\delta^4$) dependence of the Raman intensity on the detuning, which is otherwise hidden in the more accurate equation. However, it neglects the impact of spectator modes, i.e., modes that are not directly excited by the scattering event, which are still accounted for in the short-time expansion formula (eq~\ref{alpha_shorttime}) through a collective vibrational parameter $s^2$ that depends on all mode displacements.

\section{Methods}
\subsection{Computational details}
Practical implementation of the simulation protocol involves three quantum chemistry calculations for each molecule, namely the
(a) ground-state vibrational modes and frequencies (or Hessian), 
(b) electronic transition dipole moment, and 
(c) excited state gradient, all evaluated at the ground-state equilibrium geometry (i.e., Franck-Condon point). The excited-state Cartesian forces are then transformed into the normal-mode coordinates obtained from the diagonalization of the mass-scaled Hessian of the ground electronic state.
We assumed the validity of the Condon approximation, in which the transition dipole moment is a constant evaluated at a single molecular geometry, in our case the Franck-Condon point. In addition, within the DHO model, the excited state frequencies and normal modes were approximated by those of the ground state, i.e., the changes in the frequencies (mode distortion) and normal modes (Duschinsky effect) between the ground and excited electronic states were neglected.\cite{hassing1981roles} The DHO and Condon approximations were validated on the electronic absorption spectra of several dyes (see Sec.~4 of the Supporting Information). We performed the quantum-chemical calculations with the ORCA software package\cite{neese2020orca, neese2017software} and computed the Raman intensities from the aforementioned analytical expressions in a separate Python code. Further details of ab initio simulations are available in Sec. 1 of the Supporting Information.

Raman intensities were computed using mixed computational and experimental data. Namely, to avoid the computational errors in evaluating the vertical excitation gap, we estimated this energy from the experimental spectra. Since most of the probes exhibit absorption spectra that are dominated by the 0--0 transition\cite{kostjukov2022} (see Sec.~4 of the Supporting Information), we assumed that the wavelength of maximum intensity, $\lambda_{\rm max}$, corresponds to this transition. Then, within the DHO model, we could recover the experimental estimate of the vertical excitation energy as
\begin{align} \label{exp_vertical_gap}
\omega_{\rm eg, exp} = \omega_{\rm max} + \frac12 \sum_\nu \omega_\nu \Delta_\nu^2,
\end{align}
where $\omega_{\rm max}$ is the frequency of maximum absorption. Computational data was used for all other parameters, including the transition dipole moments (analyzed in Sec.~5 of the Supporting Information). After evaluating the time correlation function $C(t)$ within the DHO model (eq~\ref{eq:DHO}), we used a Gaussian dephasing $\Gamma(t) = e^{-\Theta^2 t^2 / 2 }$ in eq~\ref{eq:alpha_time} to account for the inhomogeneous broadening of the lineshape\cite{de2019efficient, petrenko2007analysis} in the solution. $\Theta$ was set to $250$ cm$^{-1}$, which was consistent with the broadening used in the electronic absorption spectra simulations.

\subsection{Experimental details}
There are two independent laser systems of different fundamental wavelengths for measuring epr-SRS signals of the molecules, providing different detuning for the same molecule. One has a fundamental wavelength of 1031.2 nm (2 ps pulse width, 80 MHz repetition rate) and the other one has a fundamental wavelength of 1064.2 nm (6 ps pulse width, 80 MHz repetition rate). The different fundamental wavelengths (used as the Stokes beam) of the two laser systems provide different pump wavelengths for the same molecule. For the Raman peak of triple bond around 2200 cm$^{-1}$, 1031.2 nm fundamental laser sets the pump wavelength to be around 840 nm and 1064.2 nm fundamental laser sets the pump wavelength to be around 860 nm. For example, the Raman peak for the nitrile bond of MARS2228 is 2228 cm$^{-1}$, thus the pump wavelength is either 838.5 nm (when the Stokes beam is 1031.2 nm) or 860.2 nm (when the Stokes beam is 1064.2 nm). Details of the two laser systems are given in Sec. 2 of the Supporting Information. A 10 mM aqueous EdU sample is measured under each laser system as a benchmark reference sample to correct the dependence of non-resonant Raman cross section on the wavelength.

\section{Results and Discussion}

\begin{table}[!thb]
\includegraphics[width=\linewidth]{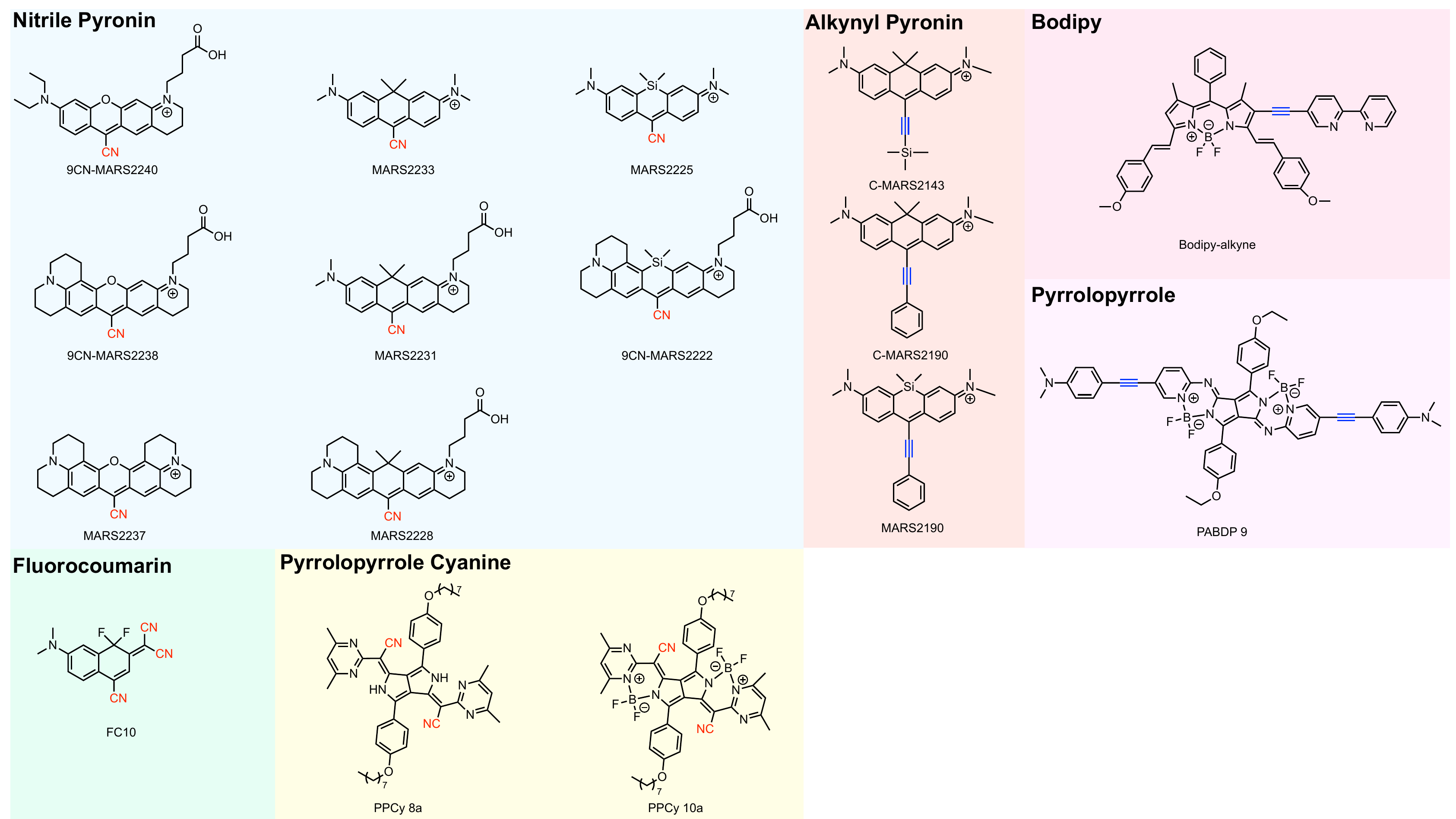}
\caption{A series of near-infrared nitrile and alkyne dyes that fit into the epr region, namely, with intense absorption in the range between 660 and 790 nm.}
\label{structures}
\end{table}

\begin{table}[!thb]
\centering
\begin{tabular}{p{0.3 \linewidth}cc}
\multicolumn{3}{l}{
Pump laser around \textbf{860} ($\pm 3$)\ nm}\\
\hline
epr-SRS probes 
& absorption maximum/nm 
& epr-SRS intensity/RIE \\
\hline
9CN-MARS2222  & 790 & 940 \\
MARS2228      & 760 & 435 \\
MARS2225      & 760 & 303 \\
MARS2231       & 744 & 240 \\
MARS2237      & 700 & 120 \\
9CN-MARS2238  & 690 & 108 \\
MARS2233      & 735 & 99 \\
9CN-MARS2240  & 675 & 86 \\
MARS2190 & 731 & 73  \\
Bodipy-alkyne & 666 & 20  \\ 
\hline
\\
\end{tabular}
\begin{tabular}{p{0.3 \linewidth}cc}
\multicolumn{3}{l}{
Pump laser around \textbf{840} ($\pm 3$)\ nm}\\
\hline
epr-SRS probes 
& absorption maximum/nm 
& epr-SRS intensity/RIE \\
\hline
MARS2228 & 760 & 620 \\
MARS2231 & 744 & 324 \\
MARS2237 & 700 & 150 \\
PADBP-9 & 699 & 186  \\
MARS2190 & 731 & 132 \\
C-MARS2190 & 696 & 16 \\
C-MARS2143 & 696 & 15 \\
FC10  & 694 & 12 \\
PPCy-8a  & 690 & 0 \\
PPCy-10a  & 692 & 0 \\
\hline \\
\end{tabular}
\caption{
Absorption maxima and epr-SRS intensities for the molecular probes at two pump laser frequencies. The absorption and Raman scattering are presented for the probes dissolved in DMSO. The Raman intensities are reported in their relative intensities versus EdU (RIE).}
\label{expintensities}
\end{table}

With the computational approach in hand, we first validate it by comparing the theoretical and experimental SRS intensities of molecules with various scaffolds. 
For the benchmark, we searched for candidates based on the criteria that the molecules strongly absorb in the epr-SRS region (660--790\ nm) and that the nitrile and alkyne groups are directly conjugated to the $\pi$-system.
Fortunately, there are several molecule scaffolds fitting into our criteria, although not many. In addition to the pyronin (O/C/Si Rhodamine) scaffold presented in MARS dyes and PPCy dyes we introduced earlier, there are other scaffolds such as coumarin\cite{matikonda2020core}, bodipy\cite{majumdar2014cyclometalated} and pyrrolopyrrole\cite{zhou2018pyrrolopyrrole} (see Table~\ref{structures}). We synthesized the molecules and measured their absorption and epr-SRS signals. The data of some of the MARS dyes were adapted from previous reports \cite{wei2017super,miao20219}. 
It is noteworthy that we report epr-SRS measurements from two independent laser systems, providing different detuning for the same molecules (see Supporting Information Sec.~2 for details). The pump wavelength is tuned to be around 840 nm or 860 nm for the triple bond and the corresponding experimental results are shown in Table~\ref{expintensities}. Each laser system used EdU as the reference substance.

\begin{figure}[!ht]
    \centering 
    \includegraphics[width=\linewidth]{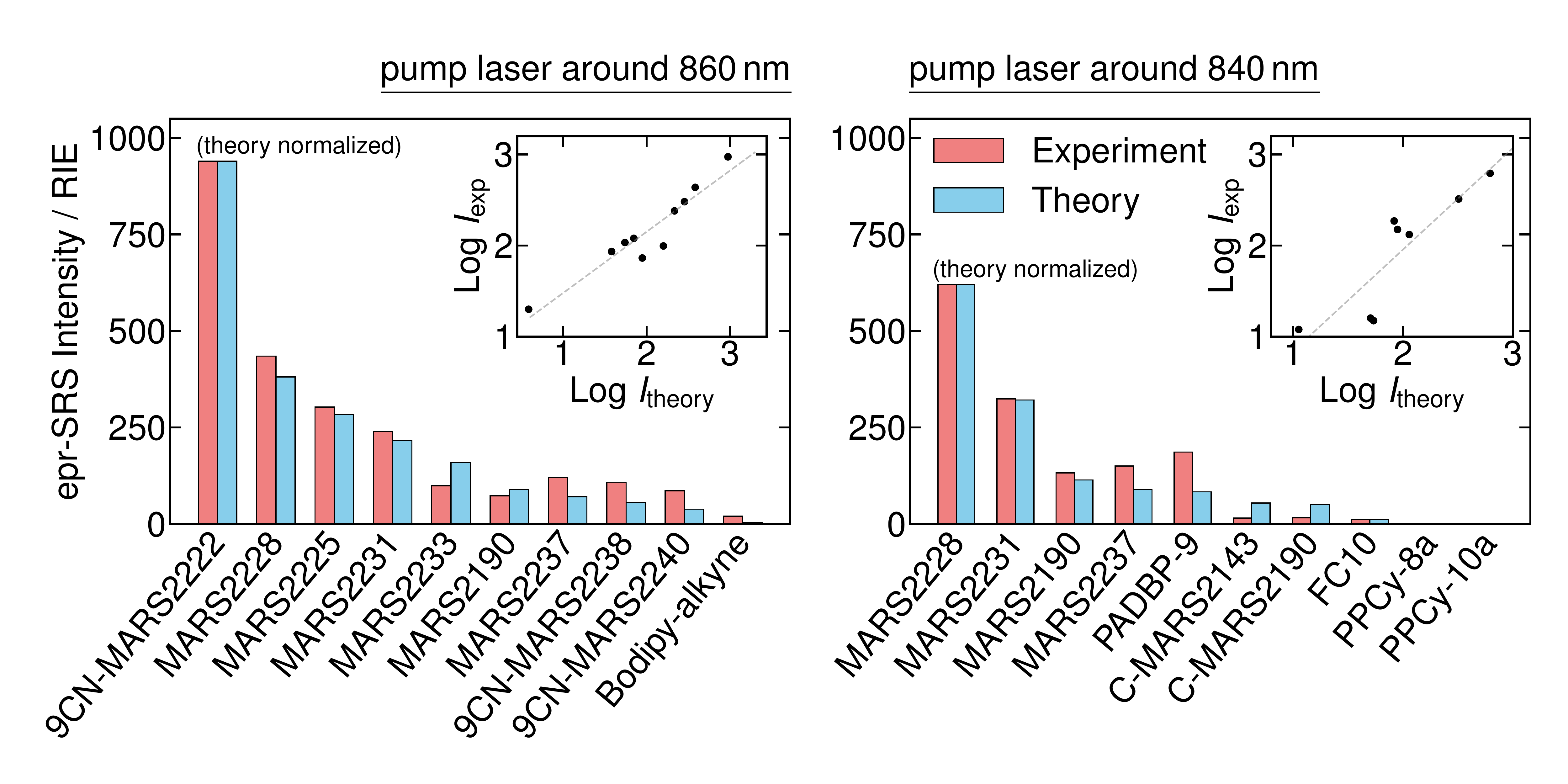}
    \caption{Comparing the measured and computed epr-SRS spectral intensities for near-infrared triple bond dyes with pump lasers at around 860 (left) and 840\,nm (right). Experimental values are reported in Table~\ref{expintensities}. The theoretical values are obtained with the DHO formula (see eqs~\ref{eq:alpha_time}, \ref{eq:DHO}-\ref{eq:delta_nu}).The insets quantitatively compare the intensities on the log$_{10}$ scale, and a linear regression (dashed gray) of the data gives the slopes of 0.67 (left, R$^2=0.95$) and 1.11 (right, R$^2=0.88$).}
    \label{fig:compare_exp_theory}
\end{figure}

Figure~\ref{fig:compare_exp_theory} compares the measured epr-SRS intensities (red bars) of nitrile and alkyne probes with the simulated intensities (blue bars) from the DHO theory described above. Data from two SRS laser systems with different pump wavelengths (around 860\,nm and 840\,nm) are shown. A consistent agreement is seen between experiment and theory across magnitudes of SRS intensities on all scaffolds. 

Furthermore, when the comparison is performed at a quantitative level in the inset, the presence of a linear trend in the figure reinforces the ability of our simple theoretical approach to identify high-intensity epr-SRS probes. A linear regression of the data gives a slope of 0.67 or 1.11 rather than 1, illustrating also the limitations of our approach. Admittedly, a number of approximations enter the simulations, including the DHO model for the time-correlation function and (TD)DFT level of electronic structure theory. Most of the calculated values (normalized to the strongest epr-SRS probe in each group) are within or close to the experimental standard error of about 10\,\% (see Figure~S1 of the Supporting Information). Yet, our calculations deviate more on the intensities of alkyne dyes: they largely overestimate the SRS intensities of the alkynyl pyronin dyes C-MARS2190 and C-MARS2143, while they underestimate the intensity of PADBP-9. Whereas in conventional simulations of a single resonance Raman spectrum the effects of detuning and transition dipole strength are only moderate, e.g., the spectra are typically scaled to the highest peak, here these factors play an important role because different molecules exhibit different detunings and transition dipole strengths. Since the epr-SRS intensity depends strongly on these parameters, even seemingly acceptable quantum-chemical errors can lead to discrepancies between theory and experiment.

\begin{figure}[!ht]
    \centering \includegraphics[width=\linewidth]{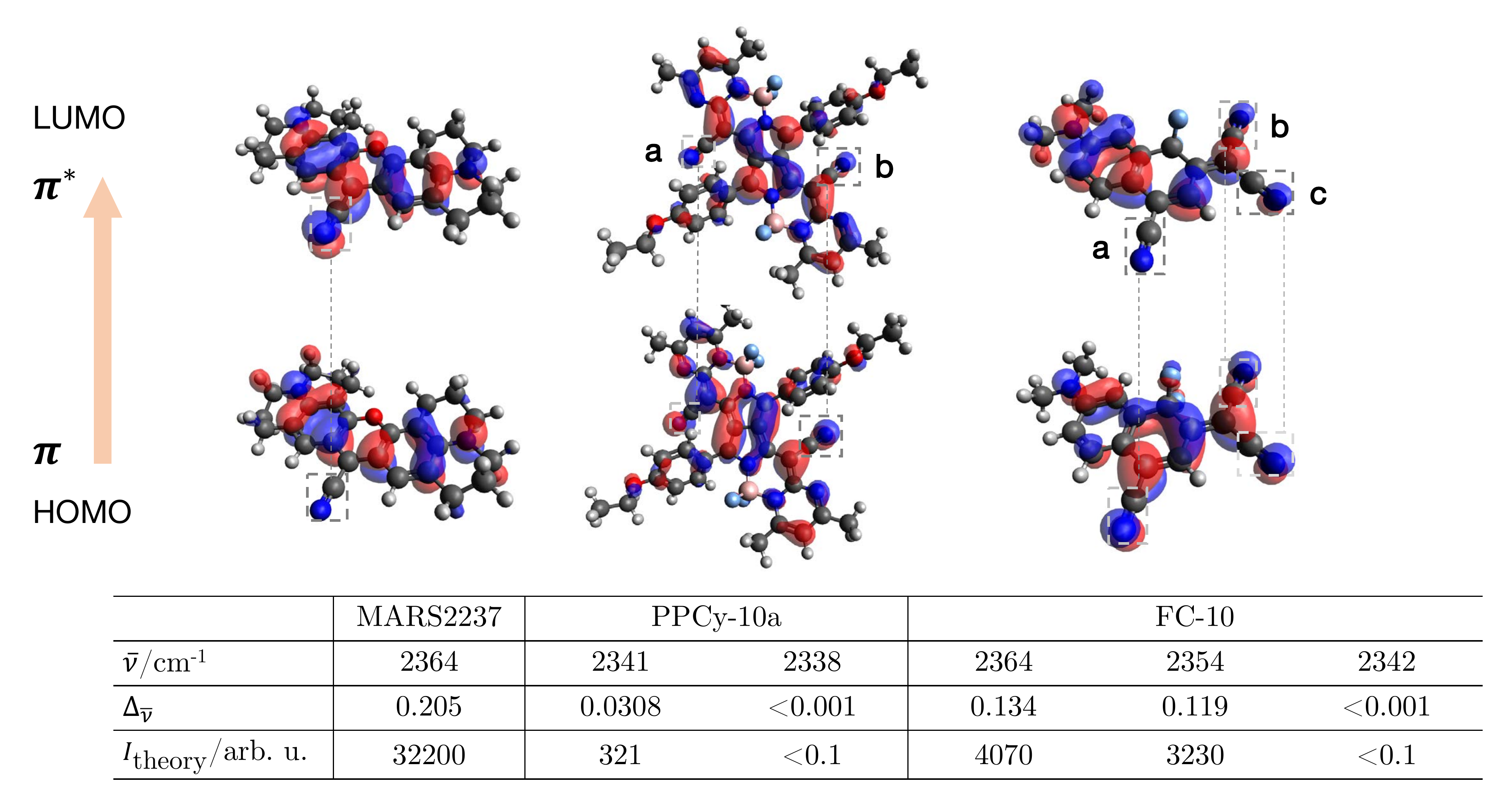}
    \caption{Epr-SRS process is strong for vibrational modes that are coupled to the electronic transition.
    $\pi \to \pi^*$ electronic transition (as indicated by the HOMO and LUMO orbitals) induces strong electron density generation on the C$\equiv$N bond for MARS2237 but not for PPCy-10a. As a result, MARS2237 exhibits significantly larger Raman mode displacement ($\Delta_{\bar{\nu}}$), hence the 100 times stronger signal intensity for the epr-SRS Raman peak associated with the C$\equiv$N vibration. The FC-10 molecule, shown in the last column exhibits two relatively strong vibrations even though it contains three nitrile groups, which can be ascribed to the symmetric (Raman active) and antisymmetric (inactive) linear combinations of the stretch vibrations corresponding to the two groups labeled b and c. }
    \label{orbital_density}
\end{figure}

More importantly, the theoretical approach allows us to analyze the effect of the vibrational mode displacement $\Delta_{\bar{\nu}}$, a key factor that enters the SRS intensity expression but cannot be easily accessed from experiments\cite{lee2019visualizing,xu2021determining}. To this end, we revisit the opening example of Figure~\ref{scheme}A. In Figure~\ref{orbital_density} we show the highest occupied molecular orbitals (HOMOs) and lowest unoccupied molecular orbitals (LUMOs) of PPCy-10a and MARS2237, as well as the corresponding parameters related to the vibrational mode displacement between the ground and excited electronic states. The $\pi \to \pi^*$ electronic transition leads to a strong electron density generation on the C$\equiv$N vibration for MARS2237 but not for PPCy-10a (Figure 3, comparing the HOMO-LUMO difference in the connected dashed-gray boxes). As we see from the $\Delta_{\bar{\nu}}$ values, the C$\equiv$N bond in MARS2237 has to stretch more to reach the equilibrium position when electronically excited, leading to a stronger SRS signal for this mode. The large difference in the epr-SRS signal strengths for the two probes then becomes straightforward according to eq~\ref{eq:DHO}, despite the fact that they exhibit similar characteristic absorption properties.

Figure~\ref{orbital_density} also reveals that simply adding more nitrile groups does not necessarily increase the overall SRS intensity. The specific comparison of MARS2237 (containing one nitrile) and PPCy-10a (containing two nitriles) is an extreme example. Additionally, in the third column of Figure~\ref{orbital_density} we present the FC-10 molecule, which contains three nitrile groups. Here, in contrast to PPCy-10a, all C$\equiv$N bonds are coupled to the electronic $\pi$ system of the dye and participate in the electronic transition. However, the coupling is weaker than in MARS2237, as demonstrated by the values of the dimensionless displacement parameter $\Delta_{\bar{\nu}}$. Interestingly, in the normal mode basis, there are only two modes with non-zero displacement. We explain this by the fact that the two C$\equiv$N bonds labeled ``b'' and ``c'' in Figure~\ref{orbital_density} are equally displaced in the excited electronic state. Therefore, their symmetric linear combination forms a normal mode that is displaced and Raman active, whereas their antisymmetric combination is not.

\begin{figure}[!t]
    \centering 
    \includegraphics[width=0.9\linewidth]{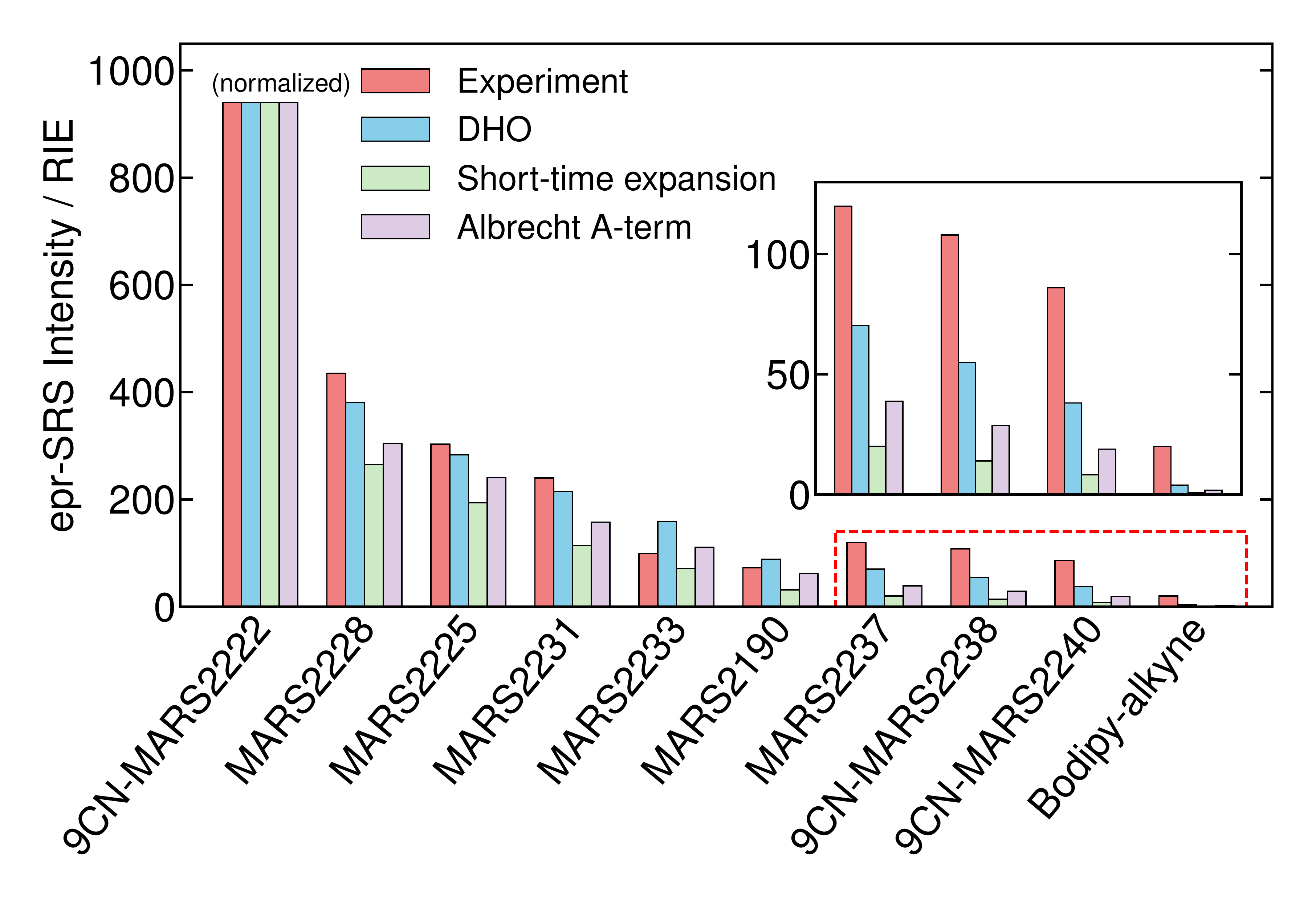}
    \caption{epr-SRS intensities simulated through the two approximate expressions---short-time expansion (eq~\ref{alpha_shorttime}) and Albrecht A-term (eq~\ref{albrecht}) formulas---compared with those simulated through the DHO equation (same as in Figure~\ref{fig:compare_exp_theory}, 860\,nm pump) and with the experiment.} 
    \label{approximate_formulas}
\end{figure}

Figure~\ref{approximate_formulas} compares the SRS intensities of several probes simulated with the full DHO expression and two approximate expressions, the short-time expansion (eq~\ref{alpha_shorttime}) and Albrecht A-term (eq~\ref{albrecht}) formulas. As expected, both expressions are less accurate than the DHO method. Since the overall computational cost of the three approaches is almost the same and is contained mostly in the required quantum chemistry calculations, the DHO expression is recommended over the other two when a quantitative agreement is needed. Nevertheless, the qualitative trends in the intensity strength are exactly reproduced. Therefore, we can use these simpler and more interpretable expressions to seek design principles for highly sensitive pre-resonance Raman probes.

Detuning ($\delta = \omega_{\rm{eg}} - \omega_{\rm{I}}$), transition dipole moment ($\mu$), and the Raman mode displacement ($\Delta_{\bar{\nu}}$) are the three factors that are highlighted in the Albrecht A-term formula (eq~\ref{albrecht}). The epr-SRS intensity increases steeply with an increase in either the transition dipole moment or the reciprocal of the detuning to the fourth order. Molecular probes with strong oscillator strengths are always preferred, while one carefully chooses the detuning, i.e. optimal pre-resonance regime, to tip the balance between high Raman signal strength and signal-to-background ratio, as discussed in the previous experimental work\cite{wei2017super}. To better illustrate the principles derived from the theoretical analysis, in Table~\ref{MARS2228_variants} we present an example where three structural isomers of (9CN-)MARS2228 are proposed and investigated as potential molecular probe candidates. The isomers, for which the nitrile group is attached at various positions on the conjugated aromatic ring, have not been reported before in the literature and are proposed here. Among the four molecules, the 9CN- substituted one exhibits the lowest-energy absorption maximum, leading to the smallest detuning and, hence, the strongest SRS intensity when the pump laser is consistently fixed at 838 nm. The 9CN- isomer also exhibits the strongest vibrational displacement and can, therefore, be expected to provide a stronger SRS signal even if the detuning is fixed to the same value for different molecules, i.e., even if the pump laser can be freely tuned for each molecule. An example of such calculations is shown in the bottom row of Table~\ref{MARS2228_variants}. Combining all of these factors together, it is not surprising that MARS2228 is listed as one of the most sensitive epr-SRS Raman probes to date. And as a pre-screening step, our computation may greatly facilitate the development of new epr-SRS scaffolds. 

We can similarly analyze the differences between the nitrile- and alkyne-based probes. For example, the alkyne MARS2190 probe exhibits a weaker SRS signal than its nitrile-based structural analog bearing a similar chromophore (Si-pyronin), the MARS2225 probe (see Table~\ref{structures} for structures and Figure~\ref{fig:compare_exp_theory} for intensities). Here, the displacement factors of around 0.146 for MARS2225 and 0.154 for MARS2190 cannot explain this discrepancy. In fact, in this case, we can explain the difference between the nitrile and alkyne dyes through their absorption properties, namely the dipole strength and detuning. Specifically, if we neglect the displacement factor, the ratio of their Raman intensities within the Albrecht approximation is $I^{\rm MARS2225}/I^{\rm MARS2190} \approx 4.12$, which agrees well with the experimental value of $303/73 \approx 4.15$ (see Table~\ref{expintensities}, pump laser at 860~nm).

The Albrecht A-term equation considers the contribution to the Raman intensity solely from the specific vibrational mode. In contrast, the short-time expansion is useful when analyzing the influence of other (spectator) vibrational modes. More specifically, eq~\ref{alpha_shorttime} can be rewritten as
\begin{align} \label{eq:multimode}
    I_{0 \to 1}^{(\bar{\nu}), \rm ST}
    = I_{0 \to 1}^{(\bar{\nu}), \text{Albrecht}}
    \left(
    \frac1{\xi^4}
    \left| \int_0^\infty t 
    e^{-t^2/2+i t/\xi} dt \right|^2 \right),
\end{align}
where the second part of the expression depends only on $\xi = s / \delta$, a dimensionless factor that involves the displacement of all modes, including spectator modes. 

When $\xi$ approaches zero, the impact of the spectator modes on the Raman intensity becomes negligible, i.e., the $\xi$-dependent term becomes $1$ (see Figure~S2), which corresponds to the relatively-large-detuning limit of Albrecht.\cite{Albrecht1961}
On the other hand, $I_{0 \to 1}^{(\bar{\nu}), \rm ST}/ I_{0 \to 1}^{(\bar{\nu}), \rm Albrecht}$ reaches its optimum at $\xi \approx 0.43 $, for which the short-time expansion gives roughly three times stronger intensity than Albrecht equation. That is to say, once the electronic transition properties associated with the Raman mode of interest are optimally tuned, additional fine functionalization of the spectator modes could further enhance the spectral intensity (see Supporting Information Sec. 3 for further discussion). However, as seen from Table~\ref{MARS2228_variants},  it is recognized as a mild effect compared with the aforementioned three key factors and is not expected to be the first target in the optimization protocol.

\begin{table}[!tb]
\renewcommand{\arraystretch}{1.25} 
\begin{tabular}{l|c|ccc}
\hline
Structure & 
\multicolumn{4}{c}{
\begin{minipage}{0.3\linewidth}
\includegraphics[width=\linewidth]{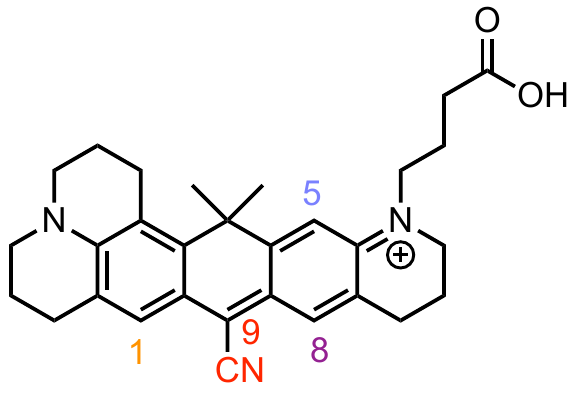}
\end{minipage}} \\ \hline
Isomers 
& {\color{red} 9CN-MARS2228}
& {\color{orange} 1CN-MARS2228} 
& {\color{blue} 5CN-MARS2228}
& {\color{violet} 8CN-MARS2228}  \\  \hline
$\lambda_{\rm sim}$/nm & 605 & 540 & 520 & 536 \\
$\mu^2$/a.u. & 37.2 & 35.8 & 30.0 & 35.7 \\
$\Delta_{\bar{\nu}}$ & 0.175 & 0.0583 & 0.0200 & 0.0595 \\
$\xi$ & 
0.41 & 
0.29 & 0.36 &
0.34  \\ \hline
$\delta$/cm$^{-1}$ & 1572 &
\multicolumn{3}{c}{\textrm{using the same detuning as 9CN-MARS2228}} \\
$I_{\rm theory}$ /arb. u. & 2.24e+05 & 
1.96e+04 &
1.81e+03 & 
2.19e+04
\end{tabular}
\caption{
Predicted epr-SRS properties for the proposed structural isomers of the molecular probe MARS2228 (labeled 9CN-MARS2228 to distinguish from other isomers). The computational protocol is kept consistent with those described above. For the calculation of epr-SRS intensities (last row), the detuning is fixed at 1572 cm$^{-1}$ (the experimental detuning of 9CN-MARS2228 with 838 nm pump laser) for all isomers. $\lambda_{\rm sim}$ denotes the vertical excitation energy obtained from the quantum chemistry simulations.
\label{MARS2228_variants}}
\end{table}

\section{Conclusion}
To conclude, we have demonstrated that theoretical modeling can elucidate the chemical principles behind vastly different epr-SRS signals from different molecular scaffolds.
The computational protocol is both robust and efficient in predicting the epr-SRS intensities and could lead towards a rational design of new epr-SRS scaffolds. Importantly, it allows us to decompose the final SRS intensity into three key factors, namely the (a) detuning, (b) transition dipole strength, and (c) vibrational mode displacement.
We showed that the strength of vibronic coupling of the specific chemical bond can be visualized by the electron density distribution during the electronic transition in the epr-SRS setting. In addition, we analyzed the applicability of approximate Albrecht and short-time expressions, which can explicitly separate these parameters. Overall, although not as accurate as the full DHO model, these approximate formulas are still useful because of their interpretability.

This work provides a fundamental step towards a computationally or data-driven methodology of epr-SRS probe design, ensuring efficient utilization of experimental efforts by avoiding traditional trial-and-error procedures. In our simulations, we have observed that the relative values of the transition dipole strength and vibrational mode displacement parameters can be sufficiently accurately modeled from quantum chemistry calculations at the TDDFT level of theory, whereas the detuning cannot. This implies two possible strategies for future work. In a purely computational approach, explicitly correlated electronic structure methods could be used for determining the vertical excitation energy to high accuracy in order to minimize the error in the detuning parameter. Alternatively, in a hybrid, experimental and computational data-driven approach, experimentally available electronic absorption spectra could be used for a range of existing dyes that do not necessarily contain a nitrile or alkyne group. Then, different isomers of such dyes with a nitrile or alkyne substitution could be computationally screened for strong vibrational mode displacement, assuming that the absorption maximum is red-shifted in a predictable way\cite{miao20219} after the addition of these functional groups. This opens new avenues for designing next-generation highly-sensitive epr-SRS palettes, driving the detection limit down to the ultimate single-molecule level. In either case, our work presents the necessary theoretical and computational basis for future design strategies.

\begin{acknowledgement}
L.W. acknowledges support from NIH Director’s New Innovator Award (GM140919). 
T.B. acknowledges financial support from the Swiss National Science Foundation through the Early Postdoc Mobility Fellowship (grant number P2ELP2-199757). 
We thank Dr. Martin J. Schnermann for sharing the FC10 dye. 
The computations presented here were conducted in the Resnick High Performance Computing Center, a facility supported by Resnick Sustainability Institute at the California Institute of Technology.
\end{acknowledgement}

\begin{suppinfo}
    Computational and experimental details, connection between short-time and Albrecht equations, Duschinsky and Herzberg--Teller effects on the absorption spectra of MARS probes, comparison between the simulated and experimental transition dipole strengths
\end{suppinfo}

\bibliography{reference.bib}

\providecommand{\latin}[1]{#1}
\makeatletter
\providecommand{\doi}
  {\begingroup\let\do\@makeother\dospecials
  \catcode`\{=1 \catcode`\}=2 \doi@aux}
\providecommand{\doi@aux}[1]{\endgroup\texttt{#1}}
\makeatother
\providecommand*\mcitethebibliography{\thebibliography}
\csname @ifundefined\endcsname{endmcitethebibliography}
  {\let\endmcitethebibliography\endthebibliography}{}
\begin{mcitethebibliography}{54}
\providecommand*\natexlab[1]{#1}
\providecommand*\mciteSetBstSublistMode[1]{}
\providecommand*\mciteSetBstMaxWidthForm[2]{}
\providecommand*\mciteBstWouldAddEndPuncttrue
  {\def\EndOfBibitem{\unskip.}}
\providecommand*\mciteBstWouldAddEndPunctfalse
  {\let\EndOfBibitem\relax}
\providecommand*\mciteSetBstMidEndSepPunct[3]{}
\providecommand*\mciteSetBstSublistLabelBeginEnd[3]{}
\providecommand*\EndOfBibitem{}
\mciteSetBstSublistMode{f}
\mciteSetBstMaxWidthForm{subitem}{(\alph{mcitesubitemcount})}
\mciteSetBstSublistLabelBeginEnd
  {\mcitemaxwidthsubitemform\space}
  {\relax}
  {\relax}

\bibitem[Ploetz \latin{et~al.}(2007)Ploetz, Laimgruber, Berner, Zinth, and
  Gilch]{ploetz2007femtosecond}
Ploetz,~E.; Laimgruber,~S.; Berner,~S.; Zinth,~W.; Gilch,~P. Femtosecond
  stimulated Raman microscopy. \emph{Appl. Phys. B} \textbf{2007}, \emph{87},
  389--393\relax
\mciteBstWouldAddEndPuncttrue
\mciteSetBstMidEndSepPunct{\mcitedefaultmidpunct}
{\mcitedefaultendpunct}{\mcitedefaultseppunct}\relax
\EndOfBibitem
\bibitem[Freudiger \latin{et~al.}(2008)Freudiger, Min, Saar, Lu, Holtom, He,
  Tsai, Kang, and Xie]{freudiger2008label}
Freudiger,~C.~W.; Min,~W.; Saar,~B.~G.; Lu,~S.; Holtom,~G.~R.; He,~C.;
  Tsai,~J.~C.; Kang,~J.~X.; Xie,~X.~S. Label-free biomedical imaging with high
  sensitivity by stimulated Raman scattering microscopy. \emph{Science}
  \textbf{2008}, \emph{322}, 1857--1861\relax
\mciteBstWouldAddEndPuncttrue
\mciteSetBstMidEndSepPunct{\mcitedefaultmidpunct}
{\mcitedefaultendpunct}{\mcitedefaultseppunct}\relax
\EndOfBibitem
\bibitem[Ozeki \latin{et~al.}(2009)Ozeki, Dake, Kajiyama, Fukui, and
  Itoh]{ozeki2009analysis}
Ozeki,~Y.; Dake,~F.; Kajiyama,~S.; Fukui,~K.; Itoh,~K. Analysis and
  experimental assessment of the sensitivity of stimulated Raman scattering
  microscopy. \emph{Opt. Express} \textbf{2009}, \emph{17}, 3651--3658\relax
\mciteBstWouldAddEndPuncttrue
\mciteSetBstMidEndSepPunct{\mcitedefaultmidpunct}
{\mcitedefaultendpunct}{\mcitedefaultseppunct}\relax
\EndOfBibitem
\bibitem[Du \latin{et~al.}(2022)Du, Wang, and Wei]{du2022bringing}
Du,~J.; Wang,~H.; Wei,~L. Bringing Vibrational Imaging to Chemical Biology with
  Molecular Probes. \emph{ACS Chem. Biol.} \textbf{2022}, \emph{17},
  1621--1637\relax
\mciteBstWouldAddEndPuncttrue
\mciteSetBstMidEndSepPunct{\mcitedefaultmidpunct}
{\mcitedefaultendpunct}{\mcitedefaultseppunct}\relax
\EndOfBibitem
\bibitem[Wei \latin{et~al.}(2017)Wei, Chen, Shi, Long, Anzalone, Zhang, Hu,
  Yuste, Cornish, and Min]{wei2017super}
Wei,~L.; Chen,~Z.; Shi,~L.; Long,~R.; Anzalone,~A.~V.; Zhang,~L.; Hu,~F.;
  Yuste,~R.; Cornish,~V.~W.; Min,~W. Super-multiplex vibrational imaging.
  \emph{Nature} \textbf{2017}, \emph{544}, 465--470\relax
\mciteBstWouldAddEndPuncttrue
\mciteSetBstMidEndSepPunct{\mcitedefaultmidpunct}
{\mcitedefaultendpunct}{\mcitedefaultseppunct}\relax
\EndOfBibitem
\bibitem[Miao \latin{et~al.}(2021)Miao, Qian, Shi, Hu, and Min]{miao20219}
Miao,~Y.; Qian,~N.; Shi,~L.; Hu,~F.; Min,~W. 9-Cyanopyronin probe palette for
  super-multiplexed vibrational imaging. \emph{Nat. Commun.} \textbf{2021},
  \emph{12}, 4518\relax
\mciteBstWouldAddEndPuncttrue
\mciteSetBstMidEndSepPunct{\mcitedefaultmidpunct}
{\mcitedefaultendpunct}{\mcitedefaultseppunct}\relax
\EndOfBibitem
\bibitem[Shi \latin{et~al.}(2022)Shi, Wei, Miao, Qian, Shi, Singer, Benninger,
  and Min]{shi2022highly}
Shi,~L.; Wei,~M.; Miao,~Y.; Qian,~N.; Shi,~L.; Singer,~R.~A.; Benninger,~R.~K.;
  Min,~W. Highly-multiplexed volumetric mapping with Raman dye imaging and
  tissue clearing. \emph{Nat. Biotechnol.} \textbf{2022}, \emph{40},
  364--373\relax
\mciteBstWouldAddEndPuncttrue
\mciteSetBstMidEndSepPunct{\mcitedefaultmidpunct}
{\mcitedefaultendpunct}{\mcitedefaultseppunct}\relax
\EndOfBibitem
\bibitem[Lee \latin{et~al.}(2021)Lee, Qian, Wang, Li, Miao, Du, Shcherbakova,
  Verkhusha, Wang, and Wei]{lee2021toward}
Lee,~D.; Qian,~C.; Wang,~H.; Li,~L.; Miao,~K.; Du,~J.; Shcherbakova,~D.~M.;
  Verkhusha,~V.~V.; Wang,~L.~V.; Wei,~L. Toward photoswitchable electronic
  pre-resonance stimulated Raman probes. \emph{J. Chem. Phys.} \textbf{2021},
  \emph{154}, 135102\relax
\mciteBstWouldAddEndPuncttrue
\mciteSetBstMidEndSepPunct{\mcitedefaultmidpunct}
{\mcitedefaultendpunct}{\mcitedefaultseppunct}\relax
\EndOfBibitem
\bibitem[Ao \latin{et~al.}(2021)Ao, Fang, Miao, Ling, Kang, Park, Wu, and
  Ji]{ao2021switchable}
Ao,~J.; Fang,~X.; Miao,~X.; Ling,~J.; Kang,~H.; Park,~S.; Wu,~C.; Ji,~M.
  Switchable stimulated Raman scattering microscopy with photochromic
  vibrational probes. \emph{Nature communications} \textbf{2021}, \emph{12},
  1--8\relax
\mciteBstWouldAddEndPuncttrue
\mciteSetBstMidEndSepPunct{\mcitedefaultmidpunct}
{\mcitedefaultendpunct}{\mcitedefaultseppunct}\relax
\EndOfBibitem
\bibitem[Shou and Ozeki(2021)Shou, and Ozeki]{shou2021photoswitchable}
Shou,~J.; Ozeki,~Y. Photoswitchable stimulated Raman scattering spectroscopy
  and microscopy. \emph{Optics Letters} \textbf{2021}, \emph{46},
  2176--2179\relax
\mciteBstWouldAddEndPuncttrue
\mciteSetBstMidEndSepPunct{\mcitedefaultmidpunct}
{\mcitedefaultendpunct}{\mcitedefaultseppunct}\relax
\EndOfBibitem
\bibitem[Fujioka \latin{et~al.}(2020)Fujioka, Shou, Kojima, Urano, Ozeki, and
  Kamiya]{fujioka2020multicolor}
Fujioka,~H.; Shou,~J.; Kojima,~R.; Urano,~Y.; Ozeki,~Y.; Kamiya,~M. Multicolor
  activatable Raman probes for simultaneous detection of plural enzyme
  activities. \emph{J. Am. Chem. Soc.} \textbf{2020}, \emph{142},
  20701--20707\relax
\mciteBstWouldAddEndPuncttrue
\mciteSetBstMidEndSepPunct{\mcitedefaultmidpunct}
{\mcitedefaultendpunct}{\mcitedefaultseppunct}\relax
\EndOfBibitem
\bibitem[Kawatani \latin{et~al.}(2022)Kawatani, Spratt, Fujioka, Shou, Misawa,
  Kojima, Urano, Ozeki, and Kamiya]{kawatani20229}
Kawatani,~M.; Spratt,~S.~J.; Fujioka,~H.; Shou,~J.; Misawa,~Y.; Kojima,~R.;
  Urano,~Y.; Ozeki,~Y.; Kamiya,~M. 9-Cyano-10-telluriumpyronin derivatives as
  red-light-activatable Raman probes. \emph{Chemistry--An Asian Journal}
  \textbf{2022}, \relax
\mciteBstWouldAddEndPunctfalse
\mciteSetBstMidEndSepPunct{\mcitedefaultmidpunct}
{}{\mcitedefaultseppunct}\relax
\EndOfBibitem
\bibitem[Xiong \latin{et~al.}(2019)Xiong, Shi, Wei, Shen, Long, Zhao, and
  Min]{xiong2019stimulated}
Xiong,~H.; Shi,~L.; Wei,~L.; Shen,~Y.; Long,~R.; Zhao,~Z.; Min,~W. Stimulated
  Raman excited fluorescence spectroscopy and imaging. \emph{Nat. photonics}
  \textbf{2019}, \emph{13}, 412--417\relax
\mciteBstWouldAddEndPuncttrue
\mciteSetBstMidEndSepPunct{\mcitedefaultmidpunct}
{\mcitedefaultendpunct}{\mcitedefaultseppunct}\relax
\EndOfBibitem
\bibitem[Fischer \latin{et~al.}(2009)Fischer, Isom{\"a}ki-Krondahl,
  G{\"o}ttker-Schnetmann, Daltrozzo, and Zumbusch]{fischer2009pyrrolopyrrole}
Fischer,~G.~M.; Isom{\"a}ki-Krondahl,~M.; G{\"o}ttker-Schnetmann,~I.;
  Daltrozzo,~E.; Zumbusch,~A. Pyrrolopyrrole cyanine dyes: A new class of
  near-infrared dyes and fluorophores. \emph{Chem. Eur. J.} \textbf{2009},
  \emph{15}, 4857--4864\relax
\mciteBstWouldAddEndPuncttrue
\mciteSetBstMidEndSepPunct{\mcitedefaultmidpunct}
{\mcitedefaultendpunct}{\mcitedefaultseppunct}\relax
\EndOfBibitem
\bibitem[Kumble \latin{et~al.}(1998)Kumble, Rush, Blackwood, Kozlowski, and
  Spiro]{Kumble1998}
Kumble,~R.; Rush,~T.~S.; Blackwood,~M.~E.; Kozlowski,~P.~M.; Spiro,~T.~G.
  {Simulation of Non-Condon Enhancement and Interference Effects in the
  Resonance Raman Intensities of Metalloporphyrins}. \emph{J. Phys. Chem. B}
  \textbf{1998}, \emph{102}, 7280--7286\relax
\mciteBstWouldAddEndPuncttrue
\mciteSetBstMidEndSepPunct{\mcitedefaultmidpunct}
{\mcitedefaultendpunct}{\mcitedefaultseppunct}\relax
\EndOfBibitem
\bibitem[Bailey \latin{et~al.}(2000)Bailey, Cohan, and
  Zink]{bailey2000interference}
Bailey,~S.~E.; Cohan,~J.~S.; Zink,~J.~I. Interference effects of multiple
  excited states in the resonance Raman spectroscopy of CpCoCOD. \emph{J. Phys.
  Chem. B} \textbf{2000}, \emph{104}, 10743--10749\relax
\mciteBstWouldAddEndPuncttrue
\mciteSetBstMidEndSepPunct{\mcitedefaultmidpunct}
{\mcitedefaultendpunct}{\mcitedefaultseppunct}\relax
\EndOfBibitem
\bibitem[Petrenko and Neese(2007)Petrenko, and Neese]{petrenko2007analysis}
Petrenko,~T.; Neese,~F. Analysis and prediction of absorption band shapes,
  fluorescence band shapes, resonance Raman intensities, and excitation
  profiles using the time-dependent theory of electronic spectroscopy. \emph{J.
  Chem. Phys.} \textbf{2007}, \emph{127}, 164319\relax
\mciteBstWouldAddEndPuncttrue
\mciteSetBstMidEndSepPunct{\mcitedefaultmidpunct}
{\mcitedefaultendpunct}{\mcitedefaultseppunct}\relax
\EndOfBibitem
\bibitem[Petrenko and Neese(2012)Petrenko, and Neese]{Petrenko2012}
Petrenko,~T.; Neese,~F. {Efficient and automatic calculation of optical band
  shapes and resonance Raman spectra for larger molecules within the
  independent mode displaced harmonic oscillator model}. \emph{J. Chem. Phys.}
  \textbf{2012}, \emph{137}, 234107\relax
\mciteBstWouldAddEndPuncttrue
\mciteSetBstMidEndSepPunct{\mcitedefaultmidpunct}
{\mcitedefaultendpunct}{\mcitedefaultseppunct}\relax
\EndOfBibitem
\bibitem[Egidi \latin{et~al.}(2014)Egidi, Bloino, Cappelli, and
  Barone]{Egidi2014}
Egidi,~F.; Bloino,~J.; Cappelli,~C.; Barone,~V. {A Robust and Effective
  Time-Independent Route to the Calculation of Resonance Raman Spectra of Large
  Molecules in Condensed Phases with the Inclusion of Duschinsky,
  Herzberg–Teller, Anharmonic, and Environmental Effects}. \emph{J. Chem.
  Theory Comput.} \textbf{2014}, \emph{10}, 346--363\relax
\mciteBstWouldAddEndPuncttrue
\mciteSetBstMidEndSepPunct{\mcitedefaultmidpunct}
{\mcitedefaultendpunct}{\mcitedefaultseppunct}\relax
\EndOfBibitem
\bibitem[Baiardi \latin{et~al.}(2014)Baiardi, Bloino, and Barone]{Baiardi2014}
Baiardi,~A.; Bloino,~J.; Barone,~V. {A general time-dependent route to
  Resonance-Raman spectroscopy including Franck-Condon, Herzberg-Teller and
  Duschinsky effects}. \emph{J. Chem. Phys.} \textbf{2014}, \emph{141},
  114108\relax
\mciteBstWouldAddEndPuncttrue
\mciteSetBstMidEndSepPunct{\mcitedefaultmidpunct}
{\mcitedefaultendpunct}{\mcitedefaultseppunct}\relax
\EndOfBibitem
\bibitem[Rao \latin{et~al.}(2016)Rao, Gelin, and Domcke]{Rao2016}
Rao,~B.~J.; Gelin,~M.~F.; Domcke,~W. {Resonant Femtosecond Stimulated Raman
  Spectra: Theory and Simulations}. \emph{J. Phys. Chem. A} \textbf{2016},
  \emph{120}, 3286--3295\relax
\mciteBstWouldAddEndPuncttrue
\mciteSetBstMidEndSepPunct{\mcitedefaultmidpunct}
{\mcitedefaultendpunct}{\mcitedefaultseppunct}\relax
\EndOfBibitem
\bibitem[Quincy \latin{et~al.}(2018)Quincy, Barclay, Caricato, and
  Elles]{Quincy2018}
Quincy,~T.~J.; Barclay,~M.~S.; Caricato,~M.; Elles,~C.~G. {Probing Dynamics in
  Higher-Lying Electronic States with Resonance-Enhanced Femtosecond Stimulated
  Raman Spectroscopy}. \emph{J. Phys. Chem. A} \textbf{2018}, \emph{122},
  8308--8319\relax
\mciteBstWouldAddEndPuncttrue
\mciteSetBstMidEndSepPunct{\mcitedefaultmidpunct}
{\mcitedefaultendpunct}{\mcitedefaultseppunct}\relax
\EndOfBibitem
\bibitem[de~Souza \latin{et~al.}(2019)de~Souza, Farias, Neese, and
  Izs{\'a}k]{de2019efficient}
de~Souza,~B.; Farias,~G.; Neese,~F.; Izs{\'a}k,~R. Efficient simulation of
  overtones and combination bands in resonant Raman spectra. \emph{J. Chem.
  Phys.} \textbf{2019}, \emph{150}, 214102\relax
\mciteBstWouldAddEndPuncttrue
\mciteSetBstMidEndSepPunct{\mcitedefaultmidpunct}
{\mcitedefaultendpunct}{\mcitedefaultseppunct}\relax
\EndOfBibitem
\bibitem[Mattiat and Luber(2021)Mattiat, and Luber]{Mattiat2021}
Mattiat,~J.; Luber,~S. {Time Domain Simulation of (Resonance) Raman Spectra of
  Liquids in the Short Time Approximation}. \emph{J. Chem. Theory Comput.}
  \textbf{2021}, \emph{17}, 344--356\relax
\mciteBstWouldAddEndPuncttrue
\mciteSetBstMidEndSepPunct{\mcitedefaultmidpunct}
{\mcitedefaultendpunct}{\mcitedefaultseppunct}\relax
\EndOfBibitem
\bibitem[Tannor(2006)]{tannor2006introduction}
Tannor,~D. \emph{Introduction to Quantum Mechanics: A Time-Dependent
  Perspective}; University Science Books, 2006; Chapter 14\relax
\mciteBstWouldAddEndPuncttrue
\mciteSetBstMidEndSepPunct{\mcitedefaultmidpunct}
{\mcitedefaultendpunct}{\mcitedefaultseppunct}\relax
\EndOfBibitem
\bibitem[Schatz and Ratner(2002)Schatz, and Ratner]{schatz2002quantum}
Schatz,~G.~C.; Ratner,~M.~A. \emph{Quantum mechanics in chemistry}; Courier
  Corporation, 2002\relax
\mciteBstWouldAddEndPuncttrue
\mciteSetBstMidEndSepPunct{\mcitedefaultmidpunct}
{\mcitedefaultendpunct}{\mcitedefaultseppunct}\relax
\EndOfBibitem
\bibitem[Kramers and Heisenberg(1925)Kramers, and
  Heisenberg]{kramers1925streuung}
Kramers,~H.~A.; Heisenberg,~W. {\"U}ber die streuung von strahlung durch atome.
  \emph{Z. Phys.} \textbf{1925}, \emph{31}, 681--708\relax
\mciteBstWouldAddEndPuncttrue
\mciteSetBstMidEndSepPunct{\mcitedefaultmidpunct}
{\mcitedefaultendpunct}{\mcitedefaultseppunct}\relax
\EndOfBibitem
\bibitem[Dirac(1927)]{dirac1927quantum}
Dirac,~P. A.~M. The quantum theory of dispersion. \emph{Proc. R. Soc. A}
  \textbf{1927}, \emph{114}, 710--728\relax
\mciteBstWouldAddEndPuncttrue
\mciteSetBstMidEndSepPunct{\mcitedefaultmidpunct}
{\mcitedefaultendpunct}{\mcitedefaultseppunct}\relax
\EndOfBibitem
\bibitem[Lee and Heller(1979)Lee, and Heller]{lee1979time}
Lee,~S.-Y.; Heller,~E.~J. Time-dependent theory of Raman scattering. \emph{J.
  Chem. Phys.} \textbf{1979}, \emph{71}, 4777--4788\relax
\mciteBstWouldAddEndPuncttrue
\mciteSetBstMidEndSepPunct{\mcitedefaultmidpunct}
{\mcitedefaultendpunct}{\mcitedefaultseppunct}\relax
\EndOfBibitem
\bibitem[Heller \latin{et~al.}(1982)Heller, Sundberg, and
  Tannor]{heller1982simple}
Heller,~E.~J.; Sundberg,~R.; Tannor,~D. Simple aspects of Raman scattering.
  \emph{J. Phys. Chem.} \textbf{1982}, \emph{86}, 1822--1833\relax
\mciteBstWouldAddEndPuncttrue
\mciteSetBstMidEndSepPunct{\mcitedefaultmidpunct}
{\mcitedefaultendpunct}{\mcitedefaultseppunct}\relax
\EndOfBibitem
\bibitem[Rohrdanz and Cina(2006)Rohrdanz, and Cina]{Rohrdanz_Cina:2006}
Rohrdanz,~M.~A.; Cina,~J.~A. {Probing intermolecular communication via lattice
  phonons with time-resolved coherent anti-Stokes Raman scattering}. \emph{Mol.
  Phys.} \textbf{2006}, \emph{104}, 1161--1178\relax
\mciteBstWouldAddEndPuncttrue
\mciteSetBstMidEndSepPunct{\mcitedefaultmidpunct}
{\mcitedefaultendpunct}{\mcitedefaultseppunct}\relax
\EndOfBibitem
\bibitem[Makhov and Shalashilin(2021)Makhov, and
  Shalashilin]{Makhov_Shalashilin:2021}
Makhov,~D.~V.; Shalashilin,~D.~V. {Simulation of the effect of vibrational
  pre-excitation on the dynamics of pyrrole Simulation of the effect of
  vibrational pre-excitation on the dynamics of pyrrole photo-dissociation}.
  \emph{J. Chem. Phys.} \textbf{2021}, \emph{154}, 104119\relax
\mciteBstWouldAddEndPuncttrue
\mciteSetBstMidEndSepPunct{\mcitedefaultmidpunct}
{\mcitedefaultendpunct}{\mcitedefaultseppunct}\relax
\EndOfBibitem
\bibitem[Worth and Lasorne(2020)Worth, and Lasorne]{Worth_Lasorne:2020}
Worth,~G.~A.; Lasorne,~B. Gaussian Wave Packets and the {DD-vMCG} Approach. In
  \emph{Quantum Chemistry and Dynamics of Excited States}; Gonz\'{a}les,~L.,
  Lindh,~R., Eds.; John Wiley \& Sons, Ltd, 2020; Chapter 13, pp 413--433\relax
\mciteBstWouldAddEndPuncttrue
\mciteSetBstMidEndSepPunct{\mcitedefaultmidpunct}
{\mcitedefaultendpunct}{\mcitedefaultseppunct}\relax
\EndOfBibitem
\bibitem[Conte and Ceotto(2020)Conte, and Ceotto]{Conte_Ceotto:2020}
Conte,~R.; Ceotto,~M. {Semiclassical Molecular Dynamics for Spectroscopic
  Calculations}. In \emph{{Quantum Chemistry and Dynamics of Excited States}};
  Gonz\'{a}les,~L., Lindh,~R., Eds.; {John Wiley \& Sons, Ltd}, 2020; Chapter
  19, pp 595--628\relax
\mciteBstWouldAddEndPuncttrue
\mciteSetBstMidEndSepPunct{\mcitedefaultmidpunct}
{\mcitedefaultendpunct}{\mcitedefaultseppunct}\relax
\EndOfBibitem
\bibitem[Bonfanti \latin{et~al.}(2018)Bonfanti, Petersen, Eisenbrandt,
  Burghardt, and Pollak]{Bonfanti_Pollak:2018}
Bonfanti,~M.; Petersen,~J.; Eisenbrandt,~P.; Burghardt,~I.; Pollak,~E.
  {Computation of the S1 S0 vibronic absorption spectrum of formaldehyde by
  variational Gaussian wavepacket and semiclassical IVR methods}. \emph{J.
  Chem. Theory Comput.} \textbf{2018}, \emph{14}, 5310--4323\relax
\mciteBstWouldAddEndPuncttrue
\mciteSetBstMidEndSepPunct{\mcitedefaultmidpunct}
{\mcitedefaultendpunct}{\mcitedefaultseppunct}\relax
\EndOfBibitem
\bibitem[Werther and Grossmann(2020)Werther, and
  Grossmann]{Werther_Grossmann:2020}
Werther,~M.; Grossmann,~F. {Apoptosis of moving nonorthogonal basis functions
  in many-particle quantum dynamics}. \emph{Phys. Rev. B} \textbf{2020},
  \emph{101}, 174315\relax
\mciteBstWouldAddEndPuncttrue
\mciteSetBstMidEndSepPunct{\mcitedefaultmidpunct}
{\mcitedefaultendpunct}{\mcitedefaultseppunct}\relax
\EndOfBibitem
\bibitem[Werther \latin{et~al.}(2021)Werther, Choudhury, and
  Grossmann]{Werther_Grossmann:2021}
Werther,~M.; Choudhury,~S.~L.; Grossmann,~F. {Coherent state based solutions of
  the time-dependent Schr{\"{o}}dinger equation: hierarchy of approximations to
  the variational principle}. \emph{Int. Rev. Phys. Chem.} \textbf{2021},
  \emph{40}, 81--125\relax
\mciteBstWouldAddEndPuncttrue
\mciteSetBstMidEndSepPunct{\mcitedefaultmidpunct}
{\mcitedefaultendpunct}{\mcitedefaultseppunct}\relax
\EndOfBibitem
\bibitem[Curchod and Mart\'{\i}nez(2018)Curchod, and
  Mart\'{\i}nez]{Curchod_Martinez:2018}
Curchod,~B. F.~E.; Mart\'{\i}nez,~T.~J. Ab Initio Nonadiabatic Quantum
  Molecular Dynamics. \emph{Chem. Rev.} \textbf{2018}, \emph{118},
  3305--3336\relax
\mciteBstWouldAddEndPuncttrue
\mciteSetBstMidEndSepPunct{\mcitedefaultmidpunct}
{\mcitedefaultendpunct}{\mcitedefaultseppunct}\relax
\EndOfBibitem
\bibitem[Prlj \latin{et~al.}(2020)Prlj, Begu{\v{s}}i{\'{c}}, Zhang, Fish,
  Wehrle, Zimmermann, Choi, Roulet, Moser, and
  Van{\'{i}}{\v{c}}ek]{Prlj_Vanicek:2020}
Prlj,~A.; Begu{\v{s}}i{\'{c}},~T.; Zhang,~Z.~T.; Fish,~G.~C.; Wehrle,~M.;
  Zimmermann,~T.; Choi,~S.; Roulet,~J.; Moser,~J.-E.; Van{\'{i}}{\v{c}}ek,~J.
  {Semiclassical Approach to Photophysics Beyond Kasha's Rule and Vibronic
  Spectroscopy Beyond the Condon Approximation. The Case of Azulene}. \emph{J.
  Chem. Theory Comput.} \textbf{2020}, \emph{16}, 2617--2626\relax
\mciteBstWouldAddEndPuncttrue
\mciteSetBstMidEndSepPunct{\mcitedefaultmidpunct}
{\mcitedefaultendpunct}{\mcitedefaultseppunct}\relax
\EndOfBibitem
\bibitem[Van{\'{i}}{\v{c}}ek and Begu{\v{s}}i{\'{c}}(2021)Van{\'{i}}{\v{c}}ek,
  and Begu{\v{s}}i{\'{c}}]{Vanicek_Begusic:2021}
Van{\'{i}}{\v{c}}ek,~J.; Begu{\v{s}}i{\'{c}},~T. {Ab Initio Semiclassical
  Evaluation of Vibrationally Resolved Electronic Spectra With Thawed
  Gaussians}. In \emph{Molecular Spectroscopy and Quantum Dynamics};
  Marquardt,~R., Quack,~M., Eds.; Elsevier, 2021; pp 199--229\relax
\mciteBstWouldAddEndPuncttrue
\mciteSetBstMidEndSepPunct{\mcitedefaultmidpunct}
{\mcitedefaultendpunct}{\mcitedefaultseppunct}\relax
\EndOfBibitem
\bibitem[Begu{\v{s}}i{\'{c}} and Van{\'{i}}{\v{c}}ek(2021)Begu{\v{s}}i{\'{c}},
  and Van{\'{i}}{\v{c}}ek]{Begusic_Vanicek:2021a}
Begu{\v{s}}i{\'{c}},~T.; Van{\'{i}}{\v{c}}ek,~J. Efficient semiclassical
  dynamics for vibronic spectroscopy beyond harmonic, Condon, and
  zero-temperature approximations. \emph{CHIMIA} \textbf{2021}, \emph{75},
  261\relax
\mciteBstWouldAddEndPuncttrue
\mciteSetBstMidEndSepPunct{\mcitedefaultmidpunct}
{\mcitedefaultendpunct}{\mcitedefaultseppunct}\relax
\EndOfBibitem
\bibitem[Begu{\v{s}}i{\'{c}} \latin{et~al.}(2022)Begu{\v{s}}i{\'{c}},
  Tapavicza, and Van{\'{i}}{\v{c}}ek]{Begusic_Vanicek:2022}
Begu{\v{s}}i{\'{c}},~T.; Tapavicza,~E.; Van{\'{i}}{\v{c}}ek,~J. {Applicability
  of the Thawed Gaussian Wavepacket Dynamics to the Calculation of Vibronic
  Spectra of Molecules with Double-Well Potential Energy Surfaces}. \emph{J.
  Chem. Theory Comput.} \textbf{2022}, \emph{18}, 3065--3074\relax
\mciteBstWouldAddEndPuncttrue
\mciteSetBstMidEndSepPunct{\mcitedefaultmidpunct}
{\mcitedefaultendpunct}{\mcitedefaultseppunct}\relax
\EndOfBibitem
\bibitem[Albrecht(1961)]{Albrecht1961}
Albrecht,~A.~C. {On the Theory of Raman Intensities}. \emph{J. Chem. Phys.}
  \textbf{1961}, \emph{34}, 1476--1484\relax
\mciteBstWouldAddEndPuncttrue
\mciteSetBstMidEndSepPunct{\mcitedefaultmidpunct}
{\mcitedefaultendpunct}{\mcitedefaultseppunct}\relax
\EndOfBibitem
\bibitem[Asher(1988)]{asher1988uv}
Asher,~S.~A. UV resonance Raman studies of molecular structure and dynamics:
  applications in physical and biophysical chemistry. \emph{Annu. Rev. Phys.
  Chem.} \textbf{1988}, \emph{39}, 537--588\relax
\mciteBstWouldAddEndPuncttrue
\mciteSetBstMidEndSepPunct{\mcitedefaultmidpunct}
{\mcitedefaultendpunct}{\mcitedefaultseppunct}\relax
\EndOfBibitem
\bibitem[Hassing and Mortensen(1981)Hassing, and Mortensen]{hassing1981roles}
Hassing,~S.; Mortensen,~O.~S. The roles of vibronic coupling and the Duschinsky
  effect in resonance Raman scattering. \emph{J. Mol. Spec.} \textbf{1981},
  \emph{87}, 1--17\relax
\mciteBstWouldAddEndPuncttrue
\mciteSetBstMidEndSepPunct{\mcitedefaultmidpunct}
{\mcitedefaultendpunct}{\mcitedefaultseppunct}\relax
\EndOfBibitem
\bibitem[Neese \latin{et~al.}(2020)Neese, Wennmohs, Becker, and
  Riplinger]{neese2020orca}
Neese,~F.; Wennmohs,~F.; Becker,~U.; Riplinger,~C. The ORCA quantum chemistry
  program package. \emph{J. Chem. Phys.} \textbf{2020}, \emph{152},
  224108\relax
\mciteBstWouldAddEndPuncttrue
\mciteSetBstMidEndSepPunct{\mcitedefaultmidpunct}
{\mcitedefaultendpunct}{\mcitedefaultseppunct}\relax
\EndOfBibitem
\bibitem[Neese(2017)]{neese2017software}
Neese,~F. Software update: the ORCA program system, version 4.0. \emph{Wiley
  Interdiscip. Rev. Comput. Mol. Sci.} \textbf{2017}, \emph{8}, 73--78\relax
\mciteBstWouldAddEndPuncttrue
\mciteSetBstMidEndSepPunct{\mcitedefaultmidpunct}
{\mcitedefaultendpunct}{\mcitedefaultseppunct}\relax
\EndOfBibitem
\bibitem[Kostjukov(2022)]{kostjukov2022}
Kostjukov,~V.~V. Excitation of rhodamine 800 in aqueous media: a theoretical
  investigation. \emph{J. Mol. Model.} \textbf{2022}, \emph{28}, 52\relax
\mciteBstWouldAddEndPuncttrue
\mciteSetBstMidEndSepPunct{\mcitedefaultmidpunct}
{\mcitedefaultendpunct}{\mcitedefaultseppunct}\relax
\EndOfBibitem
\bibitem[Matikonda \latin{et~al.}(2020)Matikonda, Ivanic, Gomez, Hammersley,
  and Schnermann]{matikonda2020core}
Matikonda,~S.~S.; Ivanic,~J.; Gomez,~M.; Hammersley,~G.; Schnermann,~M.~J. Core
  remodeling leads to long wavelength fluoro-coumarins. \emph{Chem. Sci.}
  \textbf{2020}, \emph{11}, 7302--7307\relax
\mciteBstWouldAddEndPuncttrue
\mciteSetBstMidEndSepPunct{\mcitedefaultmidpunct}
{\mcitedefaultendpunct}{\mcitedefaultseppunct}\relax
\EndOfBibitem
\bibitem[Majumdar \latin{et~al.}(2014)Majumdar, Yuan, Li, Le~Guennic, Ma,
  Zhang, Jacquemin, and Zhao]{majumdar2014cyclometalated}
Majumdar,~P.; Yuan,~X.; Li,~S.; Le~Guennic,~B.; Ma,~J.; Zhang,~C.;
  Jacquemin,~D.; Zhao,~J. Cyclometalated Ir (III) complexes with styryl-BODIPY
  ligands showing near IR absorption/emission: preparation, study of
  photophysical properties and application as photodynamic/luminescence imaging
  materials. \emph{J. Mater. Chem. B} \textbf{2014}, \emph{2}, 2838--2854\relax
\mciteBstWouldAddEndPuncttrue
\mciteSetBstMidEndSepPunct{\mcitedefaultmidpunct}
{\mcitedefaultendpunct}{\mcitedefaultseppunct}\relax
\EndOfBibitem
\bibitem[Zhou \latin{et~al.}(2018)Zhou, Ma, Gao, Wang, Lo, Wong, Xu, Kinoshita,
  and Ng]{zhou2018pyrrolopyrrole}
Zhou,~Y.; Ma,~C.; Gao,~N.; Wang,~Q.; Lo,~P.-C.; Wong,~K.~S.; Xu,~Q.-H.;
  Kinoshita,~T.; Ng,~D.~K. Pyrrolopyrrole aza boron dipyrromethene based
  two-photon fluorescent probes for subcellular imaging. \emph{J. Mater. Chem.
  B} \textbf{2018}, \emph{6}, 5570--5581\relax
\mciteBstWouldAddEndPuncttrue
\mciteSetBstMidEndSepPunct{\mcitedefaultmidpunct}
{\mcitedefaultendpunct}{\mcitedefaultseppunct}\relax
\EndOfBibitem
\bibitem[Lee \latin{et~al.}(2019)Lee, Crampton, Tallarida, and
  Apkarian]{lee2019visualizing}
Lee,~J.; Crampton,~K.~T.; Tallarida,~N.; Apkarian,~V. Visualizing vibrational
  normal modes of a single molecule with atomically confined light.
  \emph{Nature} \textbf{2019}, \emph{568}, 78--82\relax
\mciteBstWouldAddEndPuncttrue
\mciteSetBstMidEndSepPunct{\mcitedefaultmidpunct}
{\mcitedefaultendpunct}{\mcitedefaultseppunct}\relax
\EndOfBibitem
\bibitem[Xu \latin{et~al.}(2021)Xu, Zhu, Tan, Zhang, Li, Tian, Shan, Cui, Zhao,
  Dong, \latin{et~al.} others]{xu2021determining}
Xu,~J.; Zhu,~X.; Tan,~S.; Zhang,~Y.; Li,~B.; Tian,~Y.; Shan,~H.; Cui,~X.;
  Zhao,~A.; Dong,~Z. \latin{et~al.}  Determining structural and chemical
  heterogeneities of surface species at the single-bond limit. \emph{Science}
  \textbf{2021}, \emph{371}, 818--822\relax
\mciteBstWouldAddEndPuncttrue
\mciteSetBstMidEndSepPunct{\mcitedefaultmidpunct}
{\mcitedefaultendpunct}{\mcitedefaultseppunct}\relax
\EndOfBibitem
\end{mcitethebibliography}


\providecommand{\latin}[1]{#1}
\makeatletter
\providecommand{\doi}
  {\begingroup\let\do\@makeother\dospecials
  \catcode`\{=1 \catcode`\}=2 \doi@aux}
\providecommand{\doi@aux}[1]{\endgroup\texttt{#1}}
\makeatother
\providecommand*\mcitethebibliography{\thebibliography}
\csname @ifundefined\endcsname{endmcitethebibliography}
  {\let\endmcitethebibliography\endthebibliography}{}
\begin{mcitethebibliography}{14}
\providecommand*\natexlab[1]{#1}
\providecommand*\mciteSetBstSublistMode[1]{}
\providecommand*\mciteSetBstMaxWidthForm[2]{}
\providecommand*\mciteBstWouldAddEndPuncttrue
  {\def\EndOfBibitem{\unskip.}}
\providecommand*\mciteBstWouldAddEndPunctfalse
  {\let\EndOfBibitem\relax}
\providecommand*\mciteSetBstMidEndSepPunct[3]{}
\providecommand*\mciteSetBstSublistLabelBeginEnd[3]{}
\providecommand*\EndOfBibitem{}
\mciteSetBstSublistMode{f}
\mciteSetBstMaxWidthForm{subitem}{(\alph{mcitesubitemcount})}
\mciteSetBstSublistLabelBeginEnd
  {\mcitemaxwidthsubitemform\space}
  {\relax}
  {\relax}

\bibitem[Grimme \latin{et~al.}(2017)Grimme, Bannwarth, and
  Shushkov]{grimme2017robust}
Grimme,~S.; Bannwarth,~C.; Shushkov,~P. A robust and accurate tight-binding
  quantum chemical method for structures, vibrational frequencies, and
  noncovalent interactions of large molecular systems parametrized for all
  spd-block elements (Z= 1--86). \emph{J. Chem. Theory Comput.} \textbf{2017},
  \emph{13}, 1989--2009\relax
\mciteBstWouldAddEndPuncttrue
\mciteSetBstMidEndSepPunct{\mcitedefaultmidpunct}
{\mcitedefaultendpunct}{\mcitedefaultseppunct}\relax
\EndOfBibitem
\bibitem[Bannwarth \latin{et~al.}(2019)Bannwarth, Ehlert, Grimme, and
  Tight-Binding]{bannwarth2019quantum}
Bannwarth,~C.; Ehlert,~S.; Grimme,~S.; Tight-Binding,~B. P. S.-C. Quantum
  Chemical Method with Multipole Electrostatics and Density-Dependent
  Dispersion Contributions. \emph{J. Chem. Theory Comput.} \textbf{2019},
  \emph{15}, 1652--1671\relax
\mciteBstWouldAddEndPuncttrue
\mciteSetBstMidEndSepPunct{\mcitedefaultmidpunct}
{\mcitedefaultendpunct}{\mcitedefaultseppunct}\relax
\EndOfBibitem
\bibitem[Barone and Cossi(1998)Barone, and Cossi]{barone1998quantum}
Barone,~V.; Cossi,~M. Quantum calculation of molecular energies and energy
  gradients in solution by a conductor solvent model. \emph{J. Phys. Chem. A}
  \textbf{1998}, \emph{102}, 1995--2001\relax
\mciteBstWouldAddEndPuncttrue
\mciteSetBstMidEndSepPunct{\mcitedefaultmidpunct}
{\mcitedefaultendpunct}{\mcitedefaultseppunct}\relax
\EndOfBibitem
\bibitem[Petersilka \latin{et~al.}(1996)Petersilka, Gossmann, and
  Gross]{petersilka1996excitation}
Petersilka,~M.; Gossmann,~U.; Gross,~E. Excitation energies from time-dependent
  density-functional theory. \emph{Physical review letters} \textbf{1996},
  \emph{76}, 1212\relax
\mciteBstWouldAddEndPuncttrue
\mciteSetBstMidEndSepPunct{\mcitedefaultmidpunct}
{\mcitedefaultendpunct}{\mcitedefaultseppunct}\relax
\EndOfBibitem
\bibitem[Perdew(1986)]{perdew1986density}
Perdew,~J.~P. Density-functional approximation for the correlation energy of
  the inhomogeneous electron gas. \emph{Phys. Rev. B} \textbf{1986}, \emph{33},
  8822\relax
\mciteBstWouldAddEndPuncttrue
\mciteSetBstMidEndSepPunct{\mcitedefaultmidpunct}
{\mcitedefaultendpunct}{\mcitedefaultseppunct}\relax
\EndOfBibitem
\bibitem[Adamo and Barone(1999)Adamo, and Barone]{adamo1999toward}
Adamo,~C.; Barone,~V. Toward reliable density functional methods without
  adjustable parameters: The PBE0 model. \emph{J. Chem. Phys.} \textbf{1999},
  \emph{110}, 6158--6170\relax
\mciteBstWouldAddEndPuncttrue
\mciteSetBstMidEndSepPunct{\mcitedefaultmidpunct}
{\mcitedefaultendpunct}{\mcitedefaultseppunct}\relax
\EndOfBibitem
\bibitem[Zhou(2018)]{zhou2018lowest}
Zhou,~P. Why the lowest electronic excitations of rhodamines are overestimated
  by time-dependent density functional theory. \emph{Int. J. Quantum Chem.}
  \textbf{2018}, \emph{118}, e25780\relax
\mciteBstWouldAddEndPuncttrue
\mciteSetBstMidEndSepPunct{\mcitedefaultmidpunct}
{\mcitedefaultendpunct}{\mcitedefaultseppunct}\relax
\EndOfBibitem
\bibitem[Becke(1988)]{becke1988density}
Becke,~A.~D. Density-functional exchange-energy approximation with correct
  asymptotic behavior. \emph{Phys. Rev. A} \textbf{1988}, \emph{38}, 3098\relax
\mciteBstWouldAddEndPuncttrue
\mciteSetBstMidEndSepPunct{\mcitedefaultmidpunct}
{\mcitedefaultendpunct}{\mcitedefaultseppunct}\relax
\EndOfBibitem
\bibitem[Iikura \latin{et~al.}(2001)Iikura, Tsuneda, Yanai, and
  Hirao]{iikura2001long}
Iikura,~H.; Tsuneda,~T.; Yanai,~T.; Hirao,~K. A long-range correction scheme
  for generalized-gradient-approximation exchange functionals. \emph{J. Chem.
  Phys.} \textbf{2001}, \emph{115}, 3540--3544\relax
\mciteBstWouldAddEndPuncttrue
\mciteSetBstMidEndSepPunct{\mcitedefaultmidpunct}
{\mcitedefaultendpunct}{\mcitedefaultseppunct}\relax
\EndOfBibitem
\bibitem[avo()]{avogardo}
Avogadro: an open-source molecular builder and visualization tool. \\ Version
  1.93.0. http://avogadro.cc/\relax
\mciteBstWouldAddEndPuncttrue
\mciteSetBstMidEndSepPunct{\mcitedefaultmidpunct}
{\mcitedefaultendpunct}{\mcitedefaultseppunct}\relax
\EndOfBibitem
\bibitem[Hanwell \latin{et~al.}(2012)Hanwell, Curtis, Lonie, Vandermeersch,
  Zurek, and Hutchison]{hanwell2012avogadro}
Hanwell,~M.~D.; Curtis,~D.~E.; Lonie,~D.~C.; Vandermeersch,~T.; Zurek,~E.;
  Hutchison,~G.~R. Avogadro: an advanced semantic chemical editor,
  visualization, and analysis platform. \emph{J. Cheminform.} \textbf{2012},
  \emph{4}, 17\relax
\mciteBstWouldAddEndPuncttrue
\mciteSetBstMidEndSepPunct{\mcitedefaultmidpunct}
{\mcitedefaultendpunct}{\mcitedefaultseppunct}\relax
\EndOfBibitem
\bibitem[Neese \latin{et~al.}(2020)Neese, Wennmohs, Becker, and
  Riplinger]{neese2020orca}
Neese,~F.; Wennmohs,~F.; Becker,~U.; Riplinger,~C. The ORCA quantum chemistry
  program package. \emph{J. Chem. Phys.} \textbf{2020}, \emph{152},
  224108\relax
\mciteBstWouldAddEndPuncttrue
\mciteSetBstMidEndSepPunct{\mcitedefaultmidpunct}
{\mcitedefaultendpunct}{\mcitedefaultseppunct}\relax
\EndOfBibitem
\bibitem[Lewis and Maroncelli(1998)Lewis, and Maroncelli]{Lewis1998}
Lewis,~J.~E.; Maroncelli,~M. {On the (uninteresting) dependence of the
  absorption and emission transition moments of coumarin 153 on solvent}.
  \emph{Chem. Phys. Lett.} \textbf{1998}, \emph{282}, 197--203\relax
\mciteBstWouldAddEndPuncttrue
\mciteSetBstMidEndSepPunct{\mcitedefaultmidpunct}
{\mcitedefaultendpunct}{\mcitedefaultseppunct}\relax
\EndOfBibitem
\end{mcitethebibliography}

\end{document}

\section{Details of ab initio simulations}
In the quantum-chemical calculations in ORCA, we used an efficient geometry pre-optimization with a tight-binding semi-empirical potential\cite{grimme2017robust, bannwarth2019quantum} and included the molecule-solvent interaction (DMSO as solvent in our experiments) through an implicit polarizable continuum model\cite{barone1998quantum}. 
Results from density functional theory (DFT) and time-dependent density functional theory (TD-DFT)\cite{petersilka1996excitation} schemes at the hybrid functional level (PBE0/cc-pVDZ) are reported.
PBE0\cite{perdew1986density, adamo1999toward} was tested for the calculation of electronic transition energies and dipole moments of the Rhodamine dyes\cite{zhou2018lowest} and, independently, our benchmark calculations confirm that the Raman intensity predictions are insensitive on whether pure\cite{becke1988density} or range-separated\cite{iikura2001long} density functional is used instead. We also performed benchmark calculations to validate the robustness of the electronic structure calculations with respect to basis set sizes and solvent environments.
The convergence criteria of the self-consistent field steps in all reported calculations was set to `very tight', meaning that an energy change lower than 10$^{-9}$ Hartree between the last two iterations was required.
For efficiency, in PPCy-8a/PPCy-10a, we replaced the octyl groups by the ethyl groups, assuming that the absorption and Raman properties are not strongly influenced by the length of the alkyl chain.
The optimized structures and the orbitals were visualized with the Avogadro software.\cite{avogardo, hanwell2012avogadro}

\section{Experimental details}

\subsection{epr-SRS spectro-microscope setup}
\textbf{For the laser with a fundamental wavelength of 1031.2 nm:} An integrated laser (picoEMERALD, Applied Physics and Electronics, Inc.) is used as a light source for both pump and Stokes beams. It produces 2 ps pump (tunable from 770 nm – 990 nm, bandwidth 0.5 nm, spectral bandwidth $\sim 7$ cm$^{-1}$) and Stokes (also called probe, 1031.2 nm, spectral bandwidth 10 cm$^{-1}$) beams with 80MHz repetition rate. Stokes beam is modulated at 20 MHz by an internal electro-optic modulator. The spatially and temporally overlapped pump and Stokes beams are introduced into an inverted multiphoton laser scanning microscopy (FV3000, Olympus), and then focused onto the sample by a 25X water objective (XLPLN25XWMP, 1.05 N.A., Olympus). Transmitted pump and Stokes beams are collected by a high N.A. condenser lens (oil immersion, 1.4 N.A., Olympus) and pass through a bandpass filter (893/209 BrightLine, 25 mm, AVR Optics) to filter out Stokes beam. A large area (10×10 mm) Si photodiode (S3590-09, Hamamatsu) is used to measure the remaining pump beam intensity. 64 V DC voltage is used on the photodiode to increase saturation threshold and reduce response time. The output current is terminated by a 50 $\Omega$ terminator and pre-filtered by an 19.2-23.6-MHz band-pass filter (BBP-21.4+, Mini-Circuits) to reduce laser and scanning noise. The signal is then demodulated by a lock-in amplifier (SR844, Stanford Research Systems) at the modulation frequency. The in-phase X output is fed back to the Olympus IO interface box (FV30-ANALOG) of the microscope. Image acquisition speed is limited by 30 {\textmu}s time constant set for the lock-in amplifier. Laser powers are monitored through image acquisition by an internal power meter and power fluctuation are controlled within 5\,\% by the laser system. 16-bit grey scale images are acquired by Fluoview software. The epr-SRS spectra are acquired by fixing the Stokes beam at 1031.2 nm and scanning the pump beam through the designated wavelength range point by point. 10 mM aqueous EdU sample is used as a standard to give RIE (Relative Intensity to EdU) of different molecule probes. The 1 mM dye solution (DMSO) is used to acquire the epr-SRS spectra. To minimize possible photobleaching, a relatively low power (20 mW on sample for pump laser and 30 mW on sample for Stokes laser) is used.

\textbf{For the laser with a fundamental wavelength of 1064.2 nm:}  A similar laser system (picoEMERALD, Applied Physics and Electronics, Inc.) but with a different fundamental wavelength of 1064.2 nm (80 MHz repetition rate) and 6 ps pulse width was used. The intensity of the Stokes beam was modulated by a built-in EOM at 8 MHz. The other set-up is the same as the previous 1031.2 nm fundamental wavelength laser system.

\subsection{Absorption spectra measurements:}

UV-Vis absorption spectra were recorded on a Varian Cary 500 UV-Vis Spectrophotometer (Agilent). The 50 {\textmu}M dye solution (DMSO) with 1 mm cuvette is used to acquire the absorption spectra.

\subsection{Uncertainty characterization of epr-SRS measurement:}
\begin{figure}
    \centering
    \includegraphics[width=0.6 \linewidth]{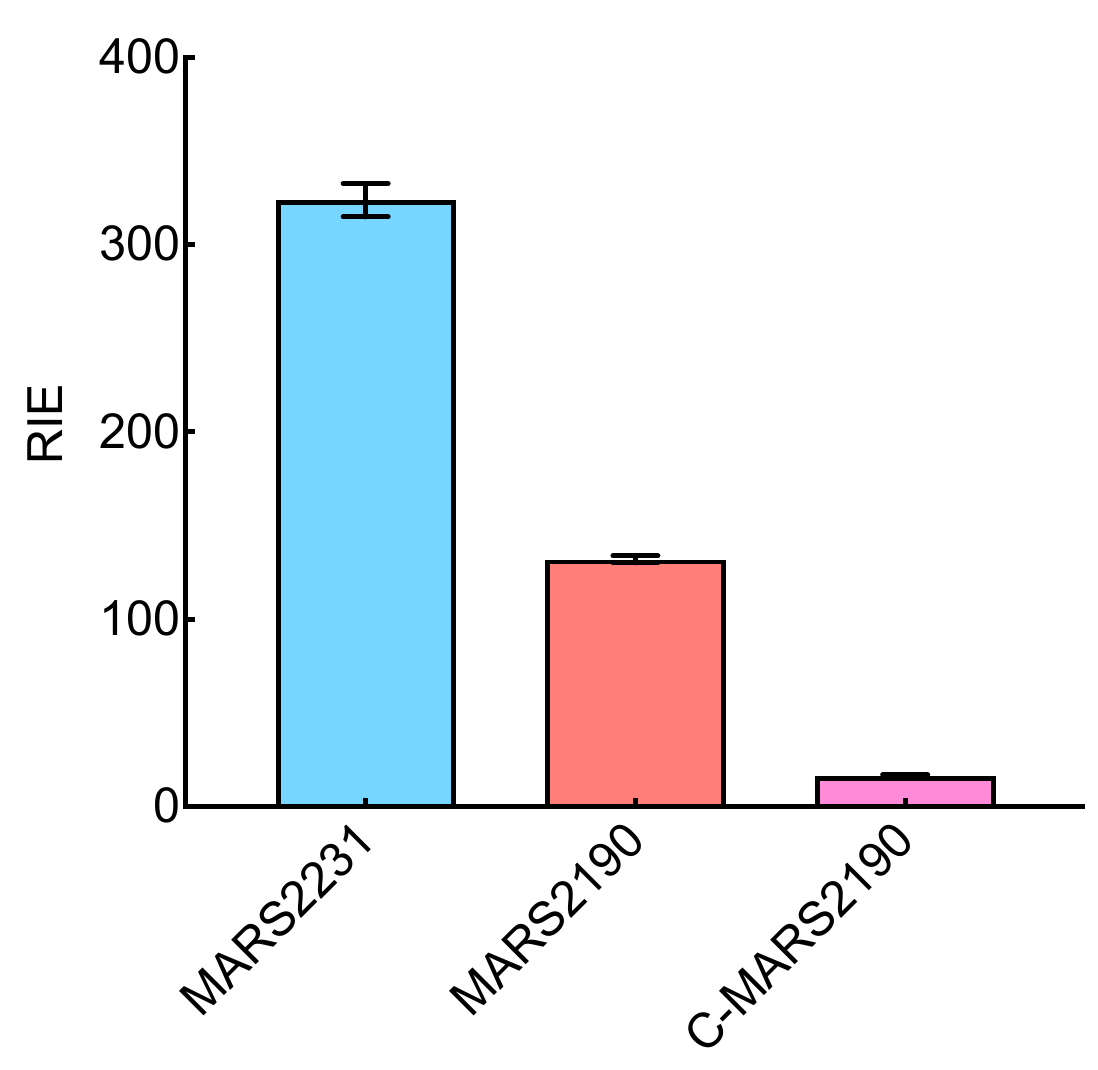}
    \caption{Uncertainty characterization of epr-SRS measurements of three dyes. Data is shown as mean $\pm$ SEM. 4--6 replicates were used for each dye.}
    \label{fig:uncertainty_analysis}
\end{figure}

To examine the uncertainty of epr-SRS measurements, we performed epr-SRS measurements of replicate samples with the 1031.2 nm fundamental laser system, we randomly chose nitrile dye MARS2231 and alkyne dye MARS2190 and C-MARS2190 for characterization, whose epr-SRS intensity spanned across different levels. The results (Figure~\ref{fig:uncertainty_analysis}) showed that the variation of each measurement is quite small: within 10\,\%.

\section{Connection between short-time and Albrecht's A-term formulas}

\begin{figure}[H]
    \centering \includegraphics[width=0.6 \linewidth]{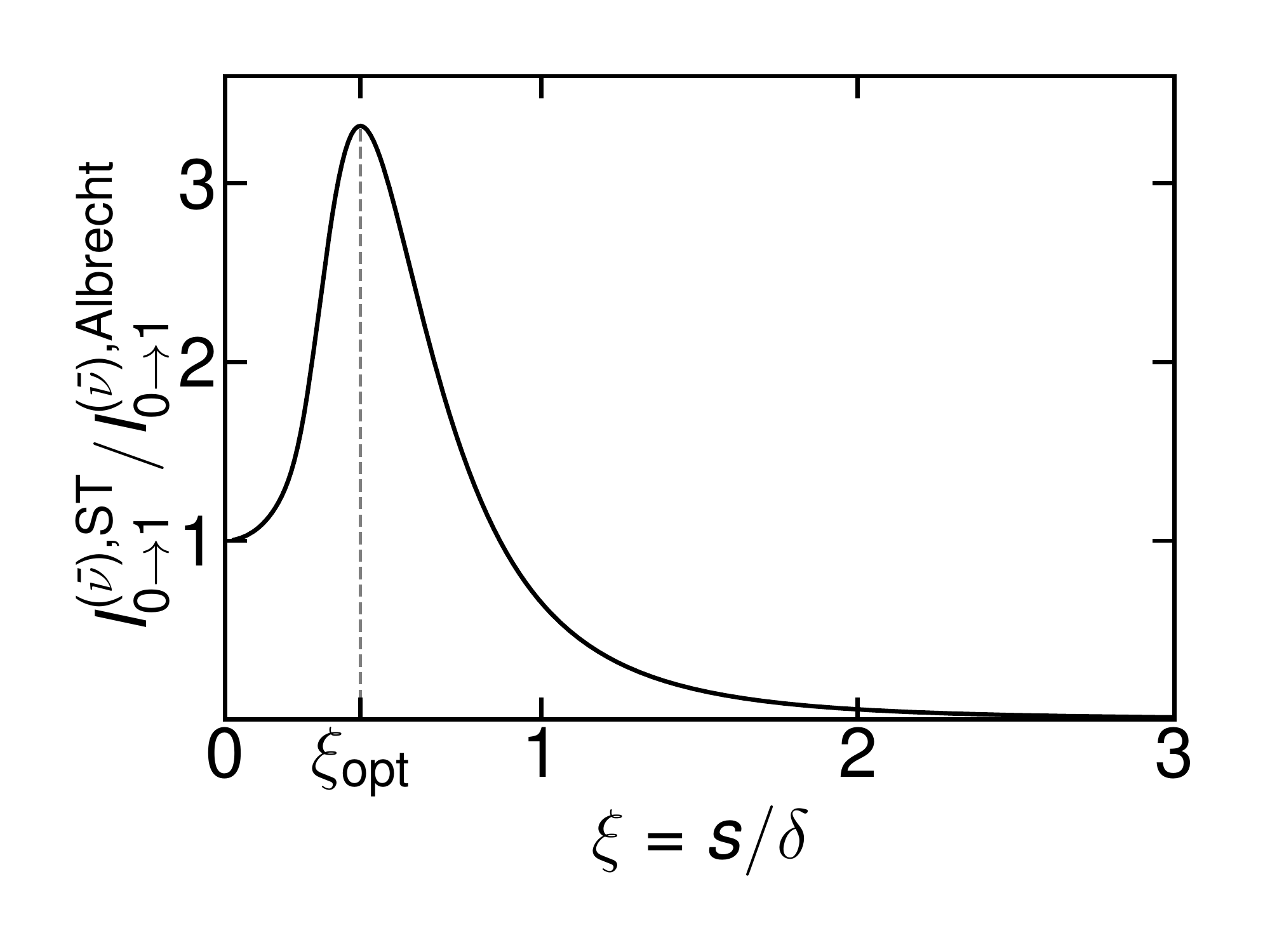}
    \caption{\label{st_albrecht_comparison} Ratio between short-time and Albrecht A-term Raman intensities as a function of $\xi = s/\delta$.}
\end{figure}

From eq~11 of the main text it follows that
\begin{equation}
    I_{0 \to 1}^{(\bar{\nu}), \rm ST}/ I_{0 \to 1}^{(\bar{\nu}), \rm Albrecht} =
    \frac{1}{\xi^4} \left|\int_0^\infty t e^{-t^2/2 + i t / \xi} dt \right|^2 = 1, \qquad \xi \rightarrow 0
\end{equation}
In other words, the Albrecht expression can be considered as a large detuning ($\delta \gg s$) limit of the short-time formula. As shown in Figure~\ref{st_albrecht_comparison}, the dependence of $I_{0 \to 1}^{(\bar{\nu}), \rm ST}/ I_{0 \to 1}^{(\bar{\nu}), \rm Albrecht}$ on $\xi$ is not monotonic. Rather, an optimal value at around $\xi_{\rm opt} \approx 0.43$ is found. However, we note that $s$, defined as $s^2 = \sum_\nu s_{\nu}^2$, depends also on the Raman mode of interest (labeled $\bar{\nu}$ in the main text) and, specifically, that $s > s_{\bar{\nu}}$. Therefore, if $\xi_{\bar{\nu}} = s_{\bar{\nu}} / \delta > \xi_{\rm opt}$, the displacements of all other modes would ideally be zero to minimize the difference $\xi - \xi_{\rm opt}$. Otherwise, if $\xi_{\bar{\nu}} < \xi_{\rm opt}$, other modes would ideally be displaced as to tune the value of $\xi$ close to the optimal one.

\section{Duschinsky and Herzberg-Teller effects}

\begin{figure}[!ht]
    \centering 
    \includegraphics[width=0.7\linewidth]{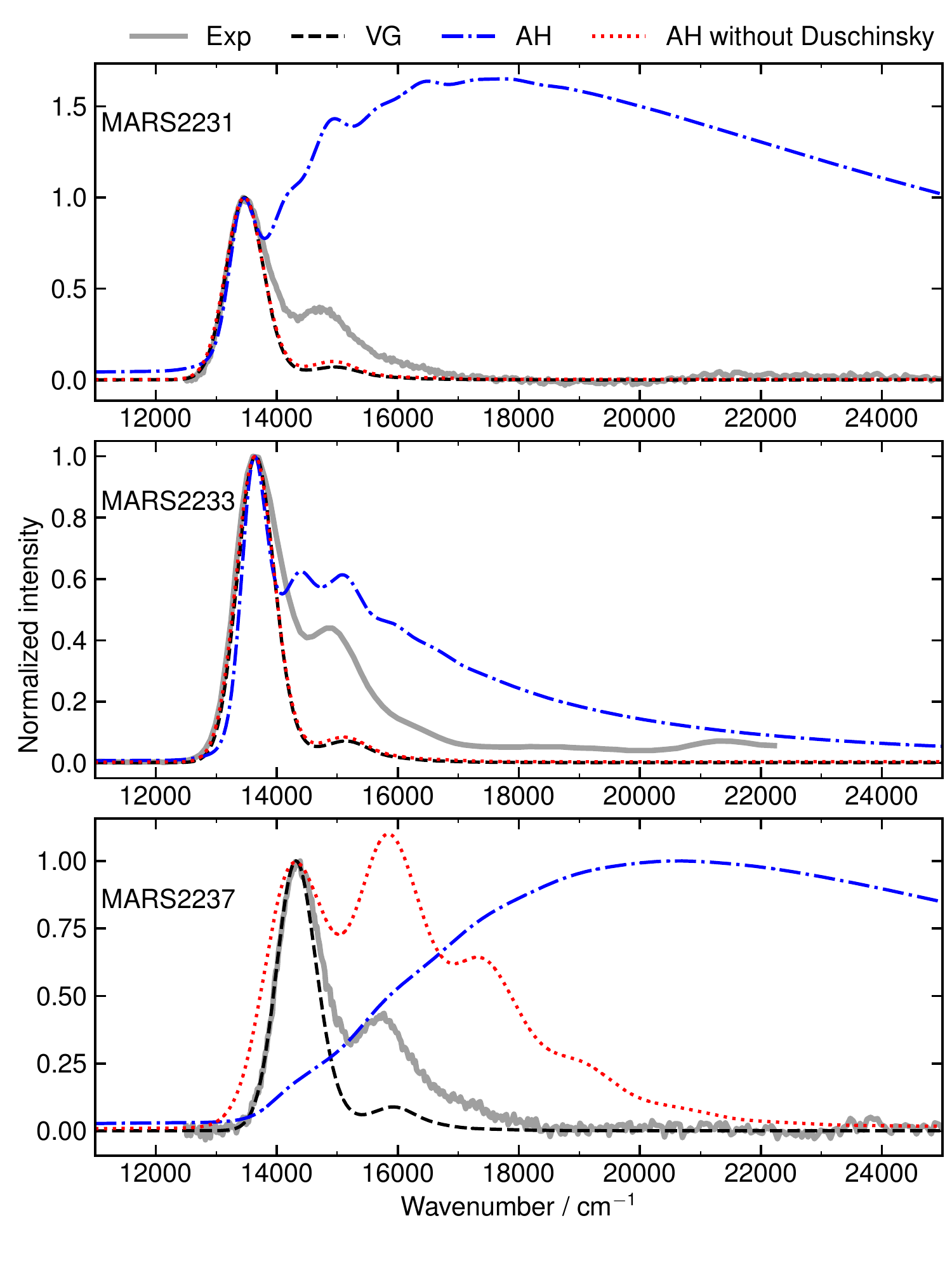}
    \caption{Experimental and simulated electronic absorption spectra of MARS2231 (top), MARS2233 (middle), MARS2237 (bottom). Simulations were based on the vertical gradient (VG) method, adiabatic Hessian (AH, including Duschinsky rotation) method, and AH method without Duschinsky rotation (by setting Duschinsky matrix $J$ to identity, option \texttt{USEJ FALSE} in Orca). All simulations were performed using the Orca \texttt{ESD} module,\cite{neese2020orca} at room temperature (298.15~K), and within the Condon approximation for the transition dipole moment. To simplify the comparison of the lineshape, the simulated spectra were scaled and shifted to match the experiment at the 0-0 transition. A Gaussian broadening function with $\sigma = 280\ $cm$^{-1}$ was used for VG and ``AH without Duschinsky'' simulations, while $\sigma = 150\ $cm$^{-1}$ was used for AH simulations.}
    \label{Duschinsky}
\end{figure}

To analyze the potential Duschinsky and Herzberg-Teller effects, we modeled the absorption spectra of the MARS2231, MARS2233, and MARS2237 probes.

First, the absorption spectra were simulated within the vertical gradient (VG) model, which is a version of the DHO model, i.e., uses the same force-constant matrix for the ground and excited electronic states, and within the adiabatic Hessian (AH) model, in which the excited-state geometry is optimized and new vibrational modes and frequencies are evaluated there. Specifically, the VG model neglects any changes in frequencies or normal-mode (Duschinsky) rotation effects, whereas the AH model fully accounts for these effects. Importantly, both of these models neglect anharmonicity of the true potential energy surfaces. Figure~\ref{Duschinsky} shows that the VG model is more reliable and robust for the studied molecules due to the incorrect overestimation of the Duschinsky coupling between the excited-state modes in the AH model (compare AH with ``AH without Duschinsky''). Although the AH approach is more accurate for MARS2233, it yields very inaccurate absorption spectra for the other two molecules. Hence, the results justify our choice of the VG model over the more expensive and less robust AH harmonic model.

\begin{figure}[!ht]
    \centering 
    \includegraphics[width=0.8\linewidth]{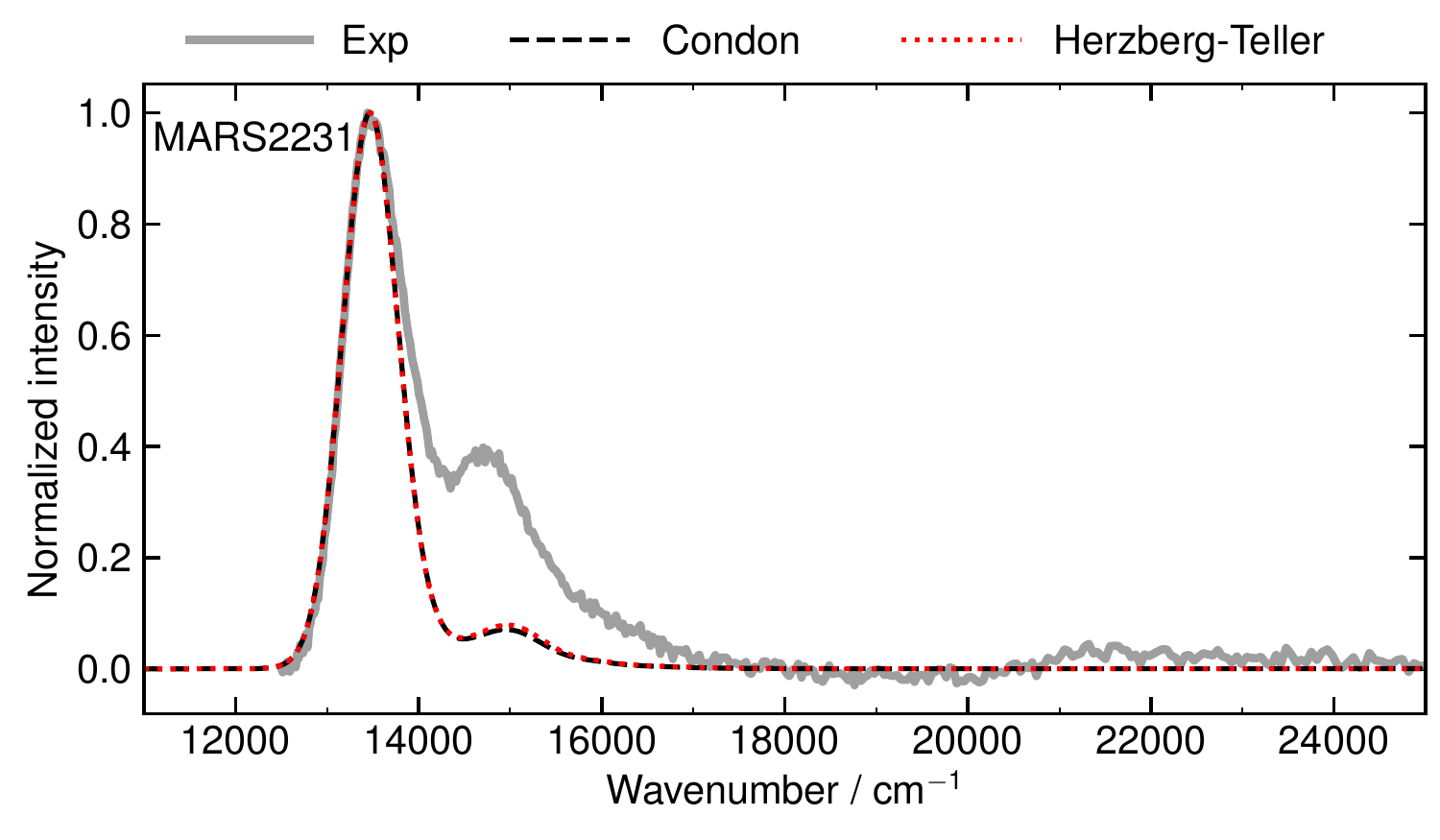}
    \caption{Experimental and simulated electronic absorption spectra of MARS2231, where the simulations were based on the VG model and used either Condon or Herzberg-Teller approximation for the transition dipole moment. See Fig.~\ref{Duschinsky} for further details.}
    \label{HT}
\end{figure}

Second, we turn to the non-Condon effects due to the dependence of the transition dipole moment on the atomic coordinates. The first-order, Herzberg-Teller correction to the Condon approximation accounts for the linear dependence of the transition dipole moment on the coordinates and is often sufficient to estimate the degree of non-Condon effects. This correction is typically necessary only in weakly allowed or symmetry-forbidden transitions. In Figure~\ref{HT}, we compare the simulations within Condon and Herzberg-Teller approximations with the experimental spectrum of MARS2231, demonstrating that the Herzberg-Teller effect is negligible.

\section{Dipole strengths: Simulation vs. experiment}
Experimental dipole strengths were determined according to\cite{Lewis1998}
\begin{equation}
    \mu_{\textrm{exp}}^2 = \frac{3 \hbar c \varepsilon_0}{\pi} \int_{-\infty}^{\infty} \frac{\sigma(\omega)}{\omega} d\omega,
\end{equation}
where the experimental absorption cross section was obtained from the molar extinction coefficient using $\sigma(\omega) = \ln(10) \varepsilon(\omega)  / N_A$.

The results (Table~\ref{oscillator_strength}) show that the simulated transition dipole moments are reliable as they only differ from the experimental values by approximately a constant factor.

\begin{table}[H]
\renewcommand{\arraystretch}{1.25}
\begin{tabular}{lcc}
\hline
epr-SRS probes & $|\mu_{\text{exp}}|$ & $|\mu_{\text{sim}}|$ \\
\hline
C-MARS2190 & 3.01 & 5.24 \\
MARS2190 & 3.26 & 5.50 \\
9CN-MARS2238 & 3.35 & 5.50 \\
9CN-MARS2240 & 3.54 & 5.42 \\
MARS2237 & 3.79 & 5.55 \\
MARS2231 & 3.89 & 5.91 \\
MARS2228 & 3.57 & 6.10 \\
PPCy-10a & 3.40 & 5.25 \\
\hline
\end{tabular}
\caption{
Magnitudes of the experimental and simulated transition dipole moments in atomic units.
}
\label{oscillator_strength}
\end{table}

\bibliography{reference.bib}